\documentclass[preprint,5p,times,twocolumn]{elsarticle}
\usepackage{amsmath,amsthm,bm}
\usepackage{amssymb}
\usepackage{color}
\usepackage{color}
\usepackage{url}
\usepackage{amsfonts}
\usepackage{numprint}
\usepackage{siunitx}
\usepackage[pagebackref=false,backref=false]{hyperref}
\usepackage{stfloats} 

\hypersetup{
  pdftitle={OPAL a Versatile Tool for Charged Particle Accelerator Simulations},
  pdfauthor={A. Adelmann et.al.},
  pdfkeywords={}
}
\usepackage{numprint}
\npthousandsep{'}
\npdecimalsign{.}
\npfourdigitnosep

\usepackage{tikz}
\usepackage{pgflibraryshapes}

\xdefinecolor{mygreen}{RGB}{0,220,0}
\xdefinecolor{myblue}{RGB}{26,150,255}

\usepackage{booktabs} 
\usepackage{algorithm}
\usepackage{algorithmic}

\usepackage{xspace}

\newcommand{\DD}{DAE$\delta$ALUS\xspace}
\newcommand{\htp}{\ensuremath{\mathrm{H}_2^+}\xspace}
\newcommand{\nuebar}{\ensuremath{\bar{\nu}_e}\xspace}

\newcommand{\spiral}{\textsc{spiral}\xspace}
\newcommand{\boxlib}{\texttt{BoxLib}}
\newcommand{\amrex}{\texttt{AMReX}}
\newcommand{\trilinos}{\texttt{Trilinos}}

\newcommand{\opal}{\textsc{OPAL}\xspace}
\newcommand{\opalt}{\textsc{OPAL-t}\xspace}
\newcommand{\opalcycl}{\textsc{OPAL-cycl}\xspace}

\newcommand{\qqref}[1]{Eq.~\eqref{#1}}
\newcommand{\figref}[1]{Fig.~\ref{#1}}
\newcommand{\tabref}[1]{Tab.~\ref{#1}}
\newcommand{\secref}[1]{\S\ref{#1}} 

\newdefinition{remark}{Remark}
\newcommand {\RM}[1]{\mathrm{#1}}

\newcommand{\V}[1]{\mathbf{#1}} 

\biboptions{sort&compress}
\graphicspath{{./figs/}}

\journal{Journal of Computational Physics (JCP)}

\begin{document}

\begin{frontmatter}

  \title{OPAL a Versatile Tool for Charged Particle Accelerator Simulations}

  \author[psi]{Andreas Adelmann\corref{cor}}
  \ead{andreas.adelmann@psi.ch}
  \author[ciemat]{Pedro Calvo}
  \author[psi]{Matthias Frey}            %
  \author[psi]{Achim Gsell}
  \author[psi]{Uldis Locans}
  \author[berlin]{Christof Metzger-Kraus}
  \author[slac]{Nicole Neveu}
  \author[ral]{Chris Rogers}
  \author[lanl]{Steve Russell}
  \author[oxford]{Suzanne Sheehy}
  \author[psi]{Jochem Snuverink}
  \author[mit]{Daniel Winklehner}

  \address[psi]{Paul Scherrer Institut, CH-5232 Villigen, Switzerland}
  \address[ciemat]{Centro de Investigaciones Energ\'eticas Medioambientales y Tecnol\'ogicas (CIEMAT), Madrid 28040, Spain}
  \address[berlin]{Open Sourcerer }
  \address[slac]{SLAC National Accelerator Laboratory, Menlo Park, CA 94025, USA}
  \address[ral]{STFC Rutherford Appleton Laboratory, Harwell Campus, Oxfordshire, OX11 0QX, UK.}
  \address[lanl]{Los Alamos National Laboratory, Los Alamos, NM 87545, USA}
  \address[oxford]{Department of Physics, University of Oxford, Oxford, OX1 3RH, UK.}
  \address[mit]{Massachusetts Institute of Technology, Cambridge, MA 02139, USA}

  \cortext[cor]{Corresponding author}

  \begin{abstract}

  Many sophisticated computer models have been developed to understand the behaviour of particle accelerators. Even these complex models often do not describe the measured data. Interactions of the beam with external fields, other particles in the same beam and the beam walls all present modelling challenges. These can be challenging to model correctly even with modern supercomputers. This paper describes \opal\ (Object Oriented Parallel Accelerator Library), a parallel open source tool for charged-particle optics in linear accelerators and rings, including 3D space charge. \opal\ is built from the ground up as a parallel application exemplifying the fact that high performance computing is the third leg of science, complementing theory and experiment.  Using the MAD language with extensions, \opal\ can run on a laptop as well as on the largest high performance computing systems. The \opal\ framework makes it easy to add new features in the form of new C++ classes, enabling the modelling of many physics processes and field types. \opal comes in two flavours: \opalcycl: tracks particles with 3D space charge including neighbouring turns in cyclotrons and FFAs with time as the independent variable. \opalt: models beam lines, linacs, rf-photo injectors and complete XFELs excluding the undulator. The code is managed through the git distributed version control system. A suite of unit tests have been developed for various parts of OPAL, validating each part of the code independently. System tests validate the overall integration of different elements.


  \end{abstract}
\end{frontmatter}

\tableofcontents

\section{Introduction}\label{sec:intro}


In recent years, precise beam dynamics simulations in the design of
high-current low-energy hadron machines as well as of fourth generation
light sources have become a very important research topic. Hadron
machines are characterised by high currents and hence require excellent
control of beam losses and/or keeping the emittance (a measure of the
phase space) of the beam in narrow ranges. This is a challenging problem
which requires the accurate modelling of the dynamics of a large ensemble
of macro or real particles subject to complicated external focusing,
accelerating and wake-fields, as well as the self-fields caused by the
Coulomb interaction of the particles.

The simulation method discussed in this paper is part of a general accelerator
modeling tool, \opal~(Object Oriented Parallel Accelerator Library)
\cite{opal}.  With \opal,  the most challenging problems in
the field of high precision particle accelerator modelling can be solved.  These problems
include the simulation of high power hadron accelerators such as Linacs, Cyclotrons, FFAs and of next
generation light sources.

Recent physics proposals include~\cite{AgarwallaConradShaevitz,daedalus,isodar}, all of them require large scale
particles based simulation in order to design and optimise the required high power hadron machines.



The paper is organised as follows, \secref{sec:arch} introduces the architecture of \opal. In \secref{sec:modelling} modelling capabilities 
followed by \secref{sec:benchm}, code benchmarks are introduced. A set of applications  presented in \secref{sec:appl}, showing the capabilities of \opal\ in ongoing
research projects. 


%
\section{Architecture of \opal}\label{sec:arch}

\subsection{Units and Conventions}

The canonical variables $\mathbf{x}$ in units of meter and $\mathbf{p}$ in units of 
$\bm{\beta}\gamma$, with $ \bm{\beta} = \mathbf{v}/c$ and $\gamma$ the relativistic factor are used. 

\subsection{The electrostatic particle-in-cell method} \label{ss:picmethod}

We consider the
Vlasov-Poisson description of the phase space, including external and
self-fields.  Let $f(\mathbf{x},\mathbf{p},t)$ be the density of the particles in the
phase space, i.e.\ the position-momentum $(\mathbf{x}, \mathbf{p})$
space.  Its evolution is determined by the collisionless \emph{Vlasov
 Poisson equation},
\begin{equation*} \label{eq:Vlasov}
  \frac{df}{dt}=\partial_t f + \mathbf{p} \cdot \nabla_{\mathbf{x}} f
  + q(\mathbf{E}+ \mathbf{p}\times\mathbf{B})\cdot
  \nabla_{\mathbf{p}} f  =  0,
\end{equation*}
where $q$ denotes the particle charge.  The
electric and magnetic fields $\mathbf{E}$ and $\mathbf{B}$ are
superpositions of external fields and self-fields (space charge),
\begin{equation}\label{eq:allfield}
    \mathbf{E} =
    \mathbf{E^{\RM{ext}}} + \mathbf{E^{\RM{self}}}, \qquad
    \mathbf{B} =
    \mathbf{B^{\RM{ext}}} + \mathbf{B^{\RM{self}}}.
\end{equation}
If $\mathbf{E}$ and $\mathbf{B}$ are known, then each particle can be
propagated according to the equation of motion for charged particles in an
electromagnetic field. After particles have moved, we have to 
update $\mathbf{E^{\RM{self}}}$ and $\mathbf{B^{\RM{self}}}$ (among other things).  
To that end, we change the coordinate system into the one moving with
the particles.  By means of the appropriate \emph{Lorentz transformation
  $\mathcal{L}$}~\cite{lali:84} we arrive at a (quasi-) static
approximation of the system in which the transformed magnetic field
becomes negligible, $\hat{\mathbf{B}}\! \approx\! \mathbf{0}$.  The
transformed electric field is then obtained from
\begin{equation}\label{eq:e-field}
  \hat{\mathbf{E}}=\hat{\mathbf{E}}^{\RM{self}}=-\nabla\hat{\phi},
\end{equation}
where the electrostatic potential $\hat{\phi}$ is the solution of the
\emph{Poisson problem}
\begin{equation}\label{eq:poisson0}
  - \Delta \hat{\phi}(\mathbf{x}) =
  \frac{ \mathcal{L}(\rho(\mathbf{x}))}{\varepsilon_0},
\end{equation}
equipped with appropriate boundary conditions.  Here, $\rho$ denotes the spatial charge
density and $\varepsilon_0$ is the dielectric constant.
By means of the inverse Lorentz transformation ( $\mathcal{L}^{-1}$) the electric field
$\hat{\mathbf{E}}$ can then be transformed back to yield both the
electric and the magnetic fields in~\eqref{eq:allfield}.

The Poisson problem~\eqref{eq:poisson0} discretized by finite
differences can efficiently be solved on a rectangular grid by a
Particle-In-Cell (PIC) approach~\cite{qiangandryne}.  The right hand side
of~\eqref{eq:poisson0} is discretized by sampling the particles at the
grid points.  In~\eqref{eq:e-field}, $\hat{\phi}$ is interpolated at the
particle positions from its values at the grid points. We also note that
the FFT-based Poisson solvers and similar
approaches~\cite{qiangandryne,qiangandgluckstern} are most effective in box-shaped or open domains.

\subsection{Equations of motion}

The flow of a particle is the solution of an initial value problem for a
differential equation, which can be approximated by numerical methods
known as (time) integrators. What is common to them is that they create
a discrete trajectory that approximates the solution of the initial
value problem. How they transport a given state from time $t_n$ to
$t_{n+1}$ is crucial for accuracy and computational effort.

In \opal\ two categories of time stepping schemes were
identified:
\renewcommand{\labelenumi}{(\alph{enumi})}
\begin{enumerate}
\item fixed step
\item geometric adaptive integration which aims at solving a regularised
  differential equation \cite{Toggweiler2014255}.
\end{enumerate}

We integrate $N$ identical particles in time, all having rest mass $m$
and charge $q$.  The relativistic equations of motion for particle $i$
are
\begin{align}
  \frac{\mathrm{d}\V{x}_i}{\mathrm{d}t} &= \frac{\V{p}_i}{m
    \gamma_i}, \label{eq:dpdt0} \\
  \frac{\mathrm{d}\V{p}_i}{\mathrm{d}t} &= q
  \left(\V{E}_i+\frac{\V{p}_i}{m \gamma_i} \times
    \V{B}_i\right), \label{eq:dpdt}
\end{align}
where $\V{x}_i$ is the position, $\V{p}_i = m \V{v}_i \gamma_i$ the
relativistic momentum, $\V{v}_i$ the velocity, $\gamma_i = 1 /
\sqrt{1-(||\V{v}_i||/c)^2} = \sqrt{1+(||\V{p}_i||/(mc))^2}$ the Lorentz
factor and $c$ the speed of light.  The electric and magnetic fields,
$\V{E}_i$ and $\V{B}_i$, can be decomposed into external field and self-field contributions:
\begin{align}
\V{E}_i &= \V{E}^{\mathrm{ext}}(\V{x}_i, t) + \V{E}^{\mathrm{self}}(i, \V{x}_{1 \ldots N}, \V{p}_{1 \ldots N}), \\
\V{B}_i &= \V{B}^{\mathrm{ext}}(\V{x}_i, t) + \V{B}^{\mathrm{self}}(i, \V{x}_{1 \ldots N}, \V{p}_{1 \ldots N}).
\end{align}
The notation $\V{x}_{1 \ldots N}$ is a shorthand for $\V{x}_1, \ldots,
\V{x}_N$, and is used for other vectors analogously. The self-field
describes the field created by the collection of particles, i.e., the
source of the Coulomb repulsion.
In this model the external electromagnetic fields (from magnets etc.),
that may explicitly depend on time, are treated independently of the particles.


\subsection{Distributions}  \label{ssec:distr}

After the description of the simulation space (i.e. the accelerator or physical system under study) is complete, the first step in actually running an \opal\ simulation is the introduction of charged particles into the problem. \opal\ can inject those particles directly into the simulation space or emit them over time in a controlled fashion (e.g. photocathode emission). How exactly this is done is directed using the \texttt{DISTRIBUTION} command.

In this section we will describe the capabilities of \opal\ for particle injection. Although we will use examples of the \texttt{DISTRIBUTION} command for illustration purposes, this is not intended as a comprehensive users guide for this feature. For more detail please see \cite{opal}.

\subsubsection{DISTRIBUTION command}
We begin with a generic example of the \texttt{DISTRIBUTION} command:
\begin{verbatim}
Name: DISTRIBUTION, TYPE = type
                    ATTRIBUTE #1 = some value,
                    ATTRIBUTE #2 = some value,
                    .
                    .
                    .
                    ATTRIBUTE #N = some value;
\end{verbatim}
This command gives the \texttt{DISTRIBUTION} a \texttt{Name}, a \texttt{TYPE} and a list of attributes that will vary depending on the value of \texttt{TYPE}. Many such \texttt{DISTRIBUTION}s can be defined in a single input with each identified by a unique name. In turn, these \texttt{DISTRIBUTION}s can be used individually, all at once, or together in sub-groups.

The first step in beam generation is the definition of a 6D particle distribution. The method \opal\ uses to do this is determined by the \texttt{DISTRIBUTION} \texttt{TYPE}, which can take on the following values:
\begin{itemize}
\item \texttt{FROMFILE}
\item \texttt{GAUSS}
\item \texttt{MATCHEDGAUSS} (See Section~\ref{sssec:matched})
\item \texttt{FLATTOP}
\item \texttt{BINOMIAL}
\end{itemize}
Types \texttt{GAUSS} to \texttt{BINOMIAL} provide options for generating traditional beam bunches. Type \texttt{MATCHEDGAUSS} is a somewhat special case of distribution generation that is described more thoroughly in Section~\ref{sssec:matched}.

Details on the capabilities of each of the \texttt{DISTRIBUTION} types can be found in \cite{opal}. However, in the case where the built in types cannot meet the user's requirements, we provide the \texttt{FROMFILE} option. This allows the user to prepare a text file describing a beam distribution that can then be read in by \opal\ and introduced to the simulation in the same way as any of the other built in options. This provides a general interface, for instance, to transfer a beam generated by another beam code into \opal\ .

\subsubsection{Beam injection} \label{sssec:beaminj}
Once a beam distribution has been defined and generated, \opal\ provides two options for introducing it into the simulation: \emph{injected} or \emph{emitted}. The \emph{injected} option is available for all flavours of \opal\ . In an \emph{injected} beam distribution, particle coordinates are described with the 6D vector,
\begin{equation*}
\mathbf{x} = (x, p_x, y, p_y, z, p_z)
\end{equation*}
At the start of the simulation, all particles appear at the same time in the simulation space.

\subsubsection{Beam emission (photocathodes)} \label{sssec:beamemis}
A second option for introducing a beam distribution into an \opal\ simulation is to emit the beam from a surface over time. The \emph{emitted} beam option is currently only available in the \opal-t\ flavour (Section~\ref{ssec:opalt}) and was implemented in order to simulate electron emission from a photocathode.

When using the \emph{emitted} option, the initial beam distribution particle coordinates are expressed differently than before, with the longitudinal position replaced by time:
\begin{equation*}
\mathbf{x} = (x, p_x, y, p_y, t, p_z)
\end{equation*}
In this case, particles are emitted gradually into the simulation space as a function of time depending on their initial time coordinate, $t$.

As already mentioned, the \emph{emitted} beam option was introduced as a way for \opal-t\ to simulate the behavior of beams from photocathodes. Of particular importance to this problem is estimating the emittance introduced to the beam by the photoemission physics, accurate calculation of the beam self-fields as the beam moves away from the cathode surface, and non-uniform electron emission. We will address the first two issues in this Section. The last will be addressed in Section~\ref{ssec:distlprof}.

The following is an example of a distribution intended to model a simple photocathode in an RF photoinjector:
\begin{verbatim}
Dist:DISTRIBUTION, TYPE          = GAUSS,
                   SIGMAX        = 0.001,
                   SIGMAY        = 0.002,
                   TRISE         = 1.0e-12,
                   TFALL         = 1.0e-12,
                   TPULSEFWHM    = 15.0e-12,
                   CUTOFFLONG    = 3.0,
                   NBIN          = 10,
                   EMISSIONSTEPS = 100,
                   EMISSIONMODEL = NONE,
                   EKIN          = 1.0,
                   EMITTED       = TRUE;
\end{verbatim}
This describes a laser pulse with an elliptical Gaussian transverse spatial profile, with a 15 ps longitudinal flat top pulse with a Gaussian rise and fall time of 1 ps. (Descriptions of the different attributes can be found in \cite{opal}.) For the discussion here, however, we are concerned with the \texttt{NBIN}, \texttt{EMISSIONSTEPS} and \texttt{EMISSIONMODEL} attributes.

As a beam is emitted from the cathode, the electrons go through very rapid acceleration, with the result that near the cathode, there can be a very large relative energy spread across the beam. This proves problematic for the electrostatic particle-in-cell method utilized by \opal\ (See Section~\ref{ss:picmethod}). A single \emph{Lorentz transformation} is not accurate in this case. However, this can be remedied by breaking the emitted beam into \texttt{NBIN} energy bins and performing \texttt{NBIN} \emph{Lorentz transformations} when computing the self-field of the beam (See Section~\ref{ssec:sc}).

Typically, when simulating a modern RF photoinjector with \opal\ , setting the integration time step to on the order of 1 ps provides accurate results. However, during the emission process, when the laser pulse is on the order of 1 - 10 ps, this time integration step proves to be too large for accurate modeling of the photoemission process. For this purpose, \opal\ provides the \texttt{EMISSIONSTEPS} parameter. This instructs the code, when emitting a beam from a photocathode, to change the integration time step temporarily to such a value so that the emission of the particles from the cathode takes \texttt{EMISSIONSTEPS} time steps. This allows for fine control over the emission calculation and more accurate results.

Finally, as RF photoinjector technology improves, accurate analysis of the thermal emittance introduced by the photocathode physics becomes more important. The \texttt{EMISSIONMODEL} option allows the user to switch between three basic models of photocathode physics (although we are hopeful that additional options will come online in the future.) These are:
\begin{itemize}
\item \texttt{EMISSIONMODEL = NONE}
\item \texttt{EMISSIONMODEL = ASTRA}
\item \texttt{EMISSIONMODEL = NONEQUIL}
\end{itemize}
The \texttt{NONE} option is as it sounds. No consideration of cathode physics is used. Instead, as particles are emitted from the photocathode surface, longitudinal momentum equivalent to the energy defined by \texttt{EKIN}, which can be set to zero, is added to each particle. This allows, for instance, the user to generate an electron beam distribution outside of \opal\ with their own photoemission model, read that distribution from file using the \texttt{FROMFILE} \texttt{DISTRIBUTION} type option and then emit it from a photocathode in an \opal\ simulation without modification.

The \texttt{ASTRA} emission model adds momentum to each particle as it is emitted from the photocathode according to:
\begin{equation*}
\begin{aligned}
p_{total} &= \sqrt{\left(\frac{EKIN}{mc^{2}} + 1\right)^{2} - 1} \\
p_{x} &= p_{total} \sin(\phi) \cos(\theta) \\
p_{y} &= p_{total} \sin(\phi) \sin(\theta) \\
p_{z} &= p_{total} |{\cos(\theta)}|
\end{aligned}
\end{equation*}
where $\theta$ is a random angle between $0$ and $\pi$, and $\phi$ is given by:
\begin{equation*}
\phi = 2.0 \arccos \left( \sqrt{x} \right)
\end{equation*}

The \texttt{NONEQUIL} is the most realistic of the photoemission models currently integrated into \opal\ . It is based on \cite{Flottmann:97}, \cite{Clendenin:00} and \cite{Dowel:09} and is valid for metal photocathodes and cathodes such as $\mathrm{CsTe}$. An example of a \texttt{DISTRIBUTION} using this model is
\begin{verbatim}
Dist:DISTRIBUTION, TYPE          = GAUSS,
                   SIGMAX        = 0.001,
                   SIGMAY        = 0.002,
                   TRISE         = 1.0e-12,
                   TFALL         = 1.0e-12,
                   TPULSEFWHM    = 15.0e-12,
                   CUTOFFLONG    = 3.0,
                   NBIN          = 10,
                   EMISSIONSTEPS = 100,
                   EMISSIONMODEL = NONEQUIL,
                   ELASER        = 6.48,
                   W             = 4.1,
                   FE            = 7.0,
                   CATHTEMP      = 325,
                   EMITTED       = TRUE;
\end{verbatim}
where \texttt{ELASER} is the drive laser energy in $eV$, \texttt{W} is the photocathode work function in $eV$, \texttt{FE} is the photocathode Fermi energy in $eV$ and \texttt{CATHTEMP} is the operating temperature of the cathode in $K$.

\subsection{Emission from a Virtual Cathode Image} \label{ssec:distlprof}

When laser images are available, i.e. taken at a
virtual cathode, they can be used to
generate a particle distribution.
This is an extension of the \texttt{FLATTOP} category,
and the definition is similar to those defined above.
The transverse radius of the bunch is still
set using the \texttt{SIGMAX} and \texttt{SIGMAY} parameters,
and the longitudinal profile is defined with
\texttt{TPULSEFWHM, TRISE}, and \texttt{TFALL}.
There are three fields specific to reading in a laser image,
\texttt{LASERPROFFN}, \texttt{INTENSITYCUT}, and \texttt{IMAGENAME}.

\begin{figure}
    \centering
    \includegraphics[width=0.45\linewidth]{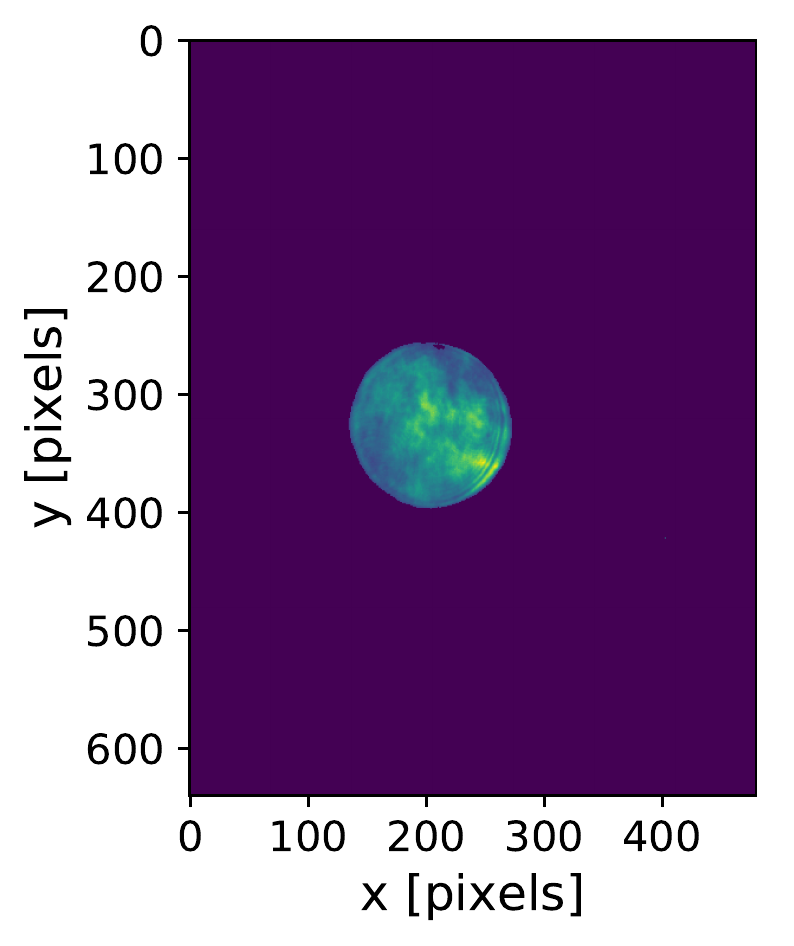}
    \includegraphics[width=0.45\linewidth]{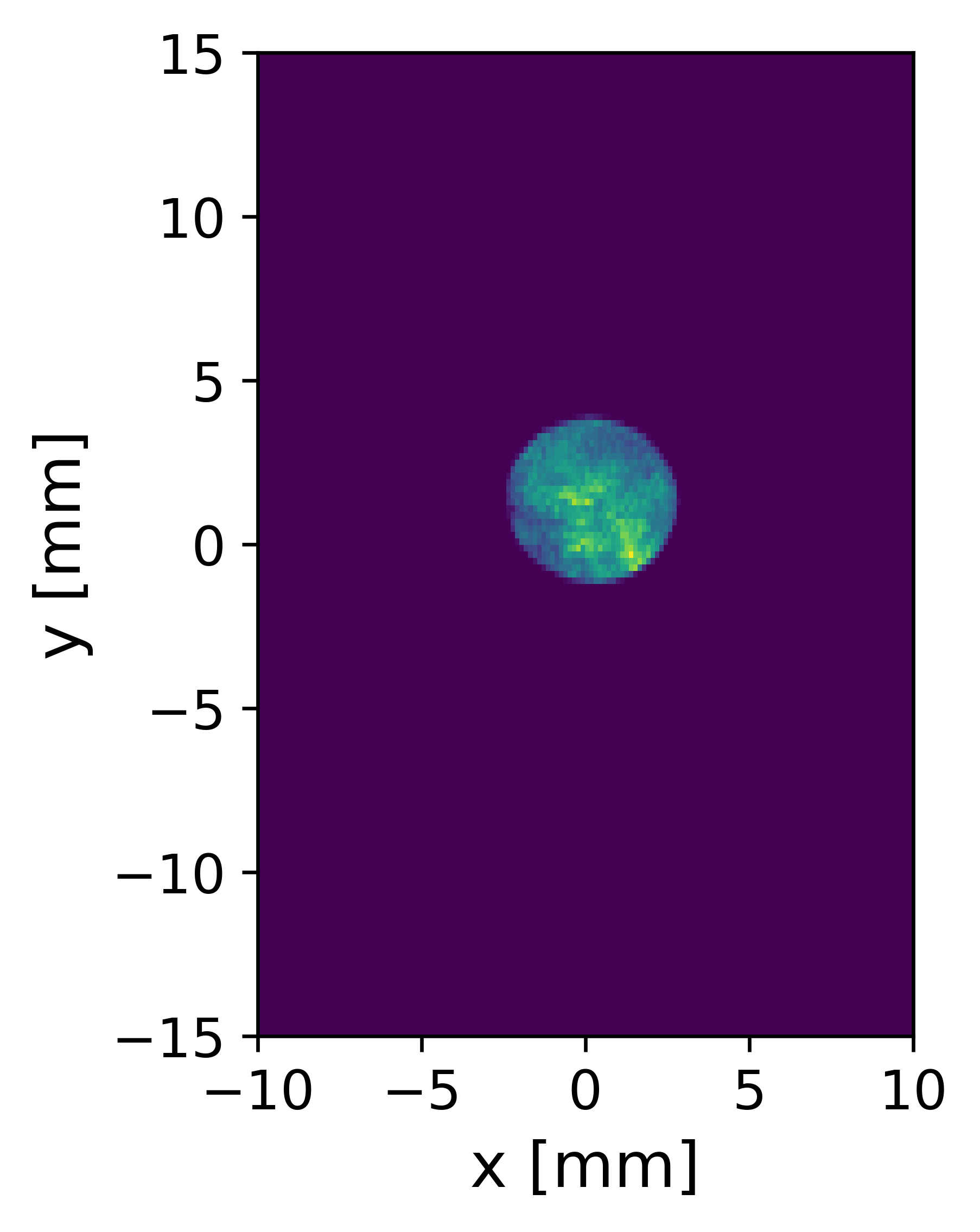}
    \caption{Comparison of virtual cathode image (left), and
    particle distribution generated by \opalt (right) using 100,000 particles.}
    \label{fig:vcc}
\end{figure}

Take the following example using a copper cathode:
\begin{verbatim}
Dist: DISTRIBUTION, TYPE = FLATTOP,
        SIGMAX = 0.0012,
        SIGMAY = 0.0012,
        TRISE = 2e-12,
        TFALL = 2e-12,
        TPULSEFWHM = 2e-12,
        CUTOFFLONG = 4.0,
        NBIN = 9,
        EMISSIONSTEPS = 100,
        EMISSIONMODEL = NONEQUIL,
        EKIN = 0.2,
        ELASER = 4.86,
        W = 4.31,
        FE = 7.0,
        CATHTEMP = 318.15,
        EMITTED = True,
        WRITETOFILE = True,
        OFFSETY = 0.0,
        LASERPROFFN = "test.h5",
        INTENSITYCUT = 0.02,
        IMAGENAME = "/Group/VCC";
\end{verbatim}
\texttt{LASERPROFFN}, is the name of the file, and
two file formats are supported; HDF5 and Portable Graymap.
\texttt{INTENSITYCUT} refers to the percentage used to subtract background.
In this example, pixels below $20\%$ of the maximum pixel intensity
will be considered background.
Finally, the \texttt{IMAGENAME}, describes how an HDF5 file is organized.
There must be a top level `Group' in the file.
The number of subcategories
and their names (`VCC' in this case) can be user defined.
Note, the center of the image is based on intensity,
and is used as the center reference (0,0).
This may not correspond to the center of the beam w.r.t. the radius,
as shown in the slight offset of Fig.~\ref{fig:vcc}.
Adjustment to the center can be made using the
\texttt{OFFSETY} and \texttt{OFFSETX}.
Similarly, these options can
be used to move the distribution further from center,
e.g. to study off axis effects in the gun.

\subsection{A truly object oriented approach}  \label{ssec:oo}
\begin{figure}[ht!]
\begin{center}
 \begin{tikzpicture}[scale=0.8, transform shape]
    \footnotesize
      \begin{scope}[shape=rectangle,rounded corners,minimum width=4.0cm,minimum height=0.6cm,fill=yellow,text centered, font=\normalsize]
       \node[fill= green!25] (0_00) at (0.1,1.5) {\textbf{MAD-Parser}};
       \node[fill= green!25] (0_00) at (3.5,1.5) {\textbf{Flavors: t,Cycl}};
       \node[fill= green!25] (0_00) at (7.0,1.5) {\textbf{Distributions}};
       \node[fill= green!25] (0_00) at (0.1,0.75) {\textbf{Sampler}};
       \node[fill= green!25] (0_00) at (3.5,0.75) {\textbf{CLASSIC}}; 
       \node[fill= green!25] (0_00) at (7.0,0.75) {\textbf{Genetic Optimizer}}; 
       \node[fill= green!25] (0_00) at (0.1,0.0) {\textbf{3D Space Charge}};
       \node[fill= green!25] (0_00) at (3.5,0.0) {\textbf{Integrators}};
       \node[fill= green!25] (0_00) at (7.0,0.0) {\textbf{PMI}};
       \node[fill= red!25] (q_00) at (0.1,-1) {\textbf{FFT}};
       \node[fill= red!25] (q_01) at (3.5,-1) {\textbf{D-Operators}};
       \node[fill= red!25] (q_02) at (7,-1) {\textbf{P-Interpolation}};
       \node[fill= red!25] (q_10) at (0.1,-1.75) {\textbf{Fields}};
       \node[fill= red!25] (q_11) at (3.5,-1.75) {\textbf{Mesh}};
       \node[fill= red!25] (q_12) at (7,-1.75) {\textbf{Particles}};
       \node[fill=red!25] (q_20) at (0.1,-2.5) {\textbf{Load Balancing}};
       \node[fill=red!25] (q_21) at (3.5,-2.5) {\textbf{Domain Decomp.}};
       \node[fill=red!25] (q_22) at (7,-2.5) {\textbf{Communication}};
       \node[fill=red!25] (q_20) at (0.1,-3.25) {\textbf{Particle-Cache}};
       \node[fill=red!25] (q_21) at (3.5,-3.25) {\textbf{PETE}};
       \node[fill=red!25] (q_22) at (7,-3.25) {\textbf{Trillions Interface}};
       \node[fill=blue!35] (q_23) at (3.5,-4.1) {\textbf{~~~~GSL --- Boost ---  H5Hut --- DKS --- NSGA-II --- Trilinos ~~~~~~}};
       \end{scope}
 \end{tikzpicture}
\end{center}
\caption{Architecture of the \opal\ framework.\ Components that are accessible via the input files are colored in green.\ In red the computer science part is denoted and in blue the external 
libraries are listed.
}
  \label{fig:over}
 \end{figure}
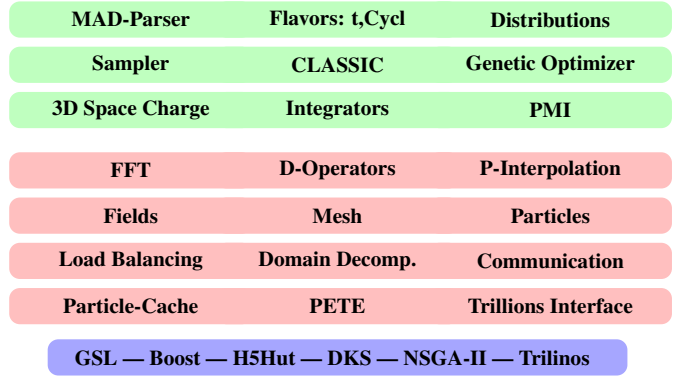


\subsection{Emerging computing architectures}  \label{ssec:emergarch}
Emerging computing architectures present the potential for large performance gains in HPC. OPAL takes advantage of emerging architectures by providing an option to offload some
compute intensive workloads to GPUs. GPU support in OPAL is built using Dynamic Kernel Scheduler (DKS) \cite{dks}, which provides a software layer between the application and the
hardware accelerator. DKS provides all the necessary functions for OPAL to handle the communication with the hardware accelerators. In general, within DKS the necessary algorithms can be built
using CUDA, OpenCL or OpenMP. However, as of now only CUDA is employed in OPAL to make use of Nvidia GPUs.

GPU support in OPAL is currently available for the FFT Poisson Solver and for particle matter interaction simulations. The FFT Poisson Solver can only by run in single node, single GPU mode. Particle matter interaction simulations can be run on multiple nodes with one GPU per node.

The performance benefits of a GPU, compared to single core CPU execution, can reach $\times11$ for the FFT Poisson solver and up to $\times120$ for the Monte Carlo simulations for particle matter interactions (tests were run with Nvidia Tesla K40 GPU) \cite{locans}. As the performance of new generations of GPUs increases, this will translate to greater performance for OPAL. We also plan to extend the OPAL DKS implementation to take advantage of computing hardware from other manufacturers.




\section{Modelling capabilities}\label{sec:modelling}
\subsection{Geometry modelling}   \label{ssec:geom}

For the simulation of precise beam dynamics, an exact modelling of the
accelerator geometry is essential. In most cases a CAD model of the
accelerator or part of it is already available. This CAD model can be
used to create a triangle mesh modelling the accelerator geometry.

Facilitating the triangle mesh \opal is able to model arbitrary
accelerator geometries and provides methods for
\begin{itemize}
\item testing whether a particle will collide with the inner surface of
the geometry (boundary) in the next time step
\item computing the distance from a given point $P$ and the boundary in
a certain direction
vector $v$ to the boundary.
\item testing whether a given point is inside the accelerator geometry
\end{itemize}

The following characteristics apply to the geometry:
\begin{itemize}
\item the geometry can consist of multiple parts
\item each part must be closed
\end{itemize}

The used methods are based on well known methods known in computer
graphics, especially ray tracing [cite]. 

\subsubsection{Initializing the geometry}
For testing whether a particle will collide with the boundary in then
next time step, we can run a line segment/triangle intersection test
for all triangle in the mesh. But even to model simple structures,
triangle meshes with thousands of elements are required. In beam
dynamics simulations, we have to run this test for all particles per
time-step, rendering the simple approach as not feasible due to
performance reasons.

In computer graphics this problem is solved by using so called voxel
meshes.  A voxel is a volume pixel representing a value on a regular
three-dimensional grid. Voxel meshes are used to render and model
three dimensional objects. Mapping a point to a voxel is a $O(1)$
operation. 

To reduce the number of required line segment/triangle intersection
tests, a voxel mesh covering the triangle mesh is created during
initialization of \opal. In a next step all triangles are assigned to
their intersecting voxels. Notes: a triangle might intersect with more
than one voxel.

For the line segment/triangle intersection tests we can now reduce the
required tests to the triangles assigned to the voxels intersecting
the line segment.

\subsubsection{Basic line segment/boundary intersection test}
\begin{itemize}
\item compute voxels intersecting the line segment/ray
\item if there are no intersecting voxels, we are done
\item run line segment/ray intersection test for all triangle associated
with intersecting voxels
\item return intersecting triangles and intersection points
\end{itemize}

\subsubsection{Particle boundary collision test}
To test whether a particle will collide with the boundary we have to
test whether the line segment given by the particle position at time t
and the expected particle position at time t+1 intersect with the
boundary.
\begin{itemize}
\item run basic line segment boundary intersection test
\item no intersecting triangles/points means no collision
\item The incidence is the closest intersection point with
respect to the particle position at time t.
\end{itemize}

\subsubsection{Compute distance from point to boundary}
Compute distance from a point $P$ inside the geometry to the boundary in a
given direction:
\begin{itemize}
\item run basic ray boundary intersection test
\item intersection point closest to P is the intersection point of ray and
boundary
\item knowing the intersection point we also know the distance from P to
the boundary
\end{itemize}

\subsubsection{Inside test}
Test whether a given point is inside the geometry:
\begin{itemize}
\item determine a reference point known to be inside or outside of the
structure. A simple solution is to choose a point outside the bounding
box of the geometry.
\item run basic line segment boundary intersection test with the reference
point and the point to test
\item if the number of intersected triangle is odd and the reference
point is outside the geometry, the testing point is inside, otherwise
outside the geometry.
\end{itemize}
Note: the reference point should preferable be close to the testing
point. This speeds up the time required to compute the intersecting
voxels.

\subsection{The \opalt flavour}  \label{ssec:opalt}

Modeling of Linacs, beam transfer lines, photoinjectors and other Linac-like accelerator systems are accomplished with the \opalt package. Calculations in \opalt are fully 3D, both for the electric and magnetic fields of the beam line objects and the beam self-fields. \opalt integrates the relativistic Lorentz equation:

\begin{equation}
    \frac{\RM{d} \gamma \vec{v}}{\RM{d} t} = \frac{q}{m}[\vec{E}_{ext} + \vec{E}_{sc} + \vec{v} \times (\vec{B}_{ext} + \vec{B}_{sc})]
\end{equation}
where $\gamma$ is the relativistic factor, $q$ is the charge, and $m$ is the rest mass of the particle. $\vec{E}$  and $\vec{B}$ are abbreviations for the electric field $\vec{E}(\vec{x},t)$ and  magnetic field $\vec{B}(\vec{x},t)$.

In \opalt the emphasis for describing accelerator systems is placed on field maps. In an \opalt model elements will typically be described by a user provided field map that is read in by the program, although there are some default maps provided for convenience (e.g. bend magnets). In this way, very accurate results are obtained as particles are integrated through those field maps, naturally taking into account higher order effects, such as fringe fields.

\opalt can read in many different field file formats that describe electrostatic, magnetostatic, and dynamic (time varying) fields. This includes 1D on axis field profiles, which are expanded off axis during beam dynamics calculations, and 2D and 3D field maps. In the latter cases \opalt uses linear interpolation to find intermediate values. For RF cavities and traveling wave structures, \opalt uses the field file to describe the electric and magnetic field profiles and multiplies them by $\sin(\omega t + \phi)$ and $\cos(\omega t + \phi)$ respectively, where $\omega$ and $\phi$ are the angular frequency and phase of the cavity.

All elements can be overlapped in an \opalt simulation. This allows for the use of a very wide range of field types and means even non-supported beam line elements can be readily used in an \opalt model as long as the user has a field map to describe it.

Beam self-fields are solved in 3D, as described in Section~\ref{ssec:sc}. This includes the ability to model image charge effects, via shifted Green's functions, and to break up a beam with large energy spread into several energy bins in order to perform individual \emph{Lorentz transformations} during a space charge calculation \cite{Qiang2006044204}. \opalt also enables the calculation of coherent synchrotron radiation effects, Section~\ref{ssec:csr}, and the inclusion of resistive wall wakefields.


\subsection{The \opalcycl Flavour}  \label{ssec:opalcyc}

The cyclotron modelling capabilities of \opal are implemented in \opalcycl \cite{OPAL-CYCL},
which solves the Vlasov-Poisson equation system in the context of cyclotron with the conditions of external static
3D magnetic field and fixed-frequency RF field. 

For high intensity isochronous cyclotrons, the space charge effects should include not only the interactions of the internal particles of a single bunch, 
but also the mutual interactions of neighboring multiple bunches in the radial direction. 
In compact AVF cyclotrons, the neighboring multi-bunch effects are particularly remarkable.
Based on the beam dynamics analysis, a ``Start-to-Stop'' model \cite{OPAL-CYCL2} and a ``Central-Bunch'' model \cite{OPAL-CYCL} are established  
for compact AVF cyclotrons with multi-turn extraction and separated-sector cyclotrons with single-turn extraction respectively. 
In both models the neighboring bunch effects are included by multi-bunch tracking. 
On that basis, the parallel Particle-In-Cell based numerical simulation algorithms are developed and implemented.
\opalcycl is developed and validated by comparing with other similar codes. 

With respect to the external magnetic field two possible situations can be considered: 
in the first situation, the measured field map data is loaded by the tracker. 
In most cases, only the vertical field, $B_z$, can be measured on the median plane ($z=0$) by using measurement equipment.
Since the magnetic field outside the median plane is required to compute trajectories with $z \neq 0$, the field needs to be expanded in the $Z$ direction. 
According to the approach given by Gordon and Taivassalo \cite{Gordon:85}, 
by using a magnetic potential and measured $B_z$ on the median plane
at the point $(r,\theta, z)$ in cylindrical polar coordinates, the third order field can be written as    
\begin{eqnarray}\label{eq:Bfield}
  B_r(r,\theta, z) & = & z\frac{\partial B_z}{\partial r}-\frac{1}{6}z^3 C_r, \nonumber\\    
  B_\theta(r,\theta, z) & = & \frac{z}{r}\frac{\partial B_z}{\partial \theta}-\frac{1}{6}\frac{z^3}{r} C_{\theta}, \\     
  B_z(r,\theta, z) & = & B_z-\frac{1}{2}z^2 C_z,  \nonumber    
\end{eqnarray}
where $B_z\equiv B_z(r, \theta, 0)$ and 
\begin{flalign}\label{eq:Bcoeff}
  C_r & = & \frac{\partial^3B_z}{\partial r^3} + \frac{1}{r}\frac{\partial^2 B_z}{\partial r^2} - \frac{1}{r^2}\frac{\partial B_z}{\partial r} 
        + \frac{1}{r^2}\frac{\partial^3 B_z}{\partial r \partial \theta^2} - 2\frac{1}{r^3}\frac{\partial^2 B_z}{\partial \theta^2}, \nonumber  \\    
  C_{\theta} & = & \frac{1}{r}\frac{\partial^2 B_z}{\partial r \partial \theta} + \frac{\partial^3 B_z}{\partial r^2 \partial \theta}
        + \frac{1}{r^2}\frac{\partial^3 B_z}{\partial \theta^3},  \\
  C_z & = & \frac{1}{r}\frac{\partial B_z}{\partial r} + \frac{\partial^2 B_z}{\partial r^2} + \frac{1}{r^2}\frac{\partial^2 B_z}{\partial \theta^2}. \nonumber
\end{flalign}
All the partial differential coefficients are computed on the median plane data by interpolation, using Lagrange's 5-point formula.

In the other situation, 3D magnetic field data for the region of interest is calculated numerically by building a 3D model using commercial software 
during the design phase of a new cyclotron. In this case the calculated field will be more accurate, especially at large distances from the median plane i.e. a
full 3D field map can be calculated. 

Finally both the external fields and space charge fields are used to track particles. 
In the current version, three different integrators are implemented: a fourth-order Runge-Kutta integrator,
a second-order Leap-Frog integrator and a multiple-time-stepping variant of the Boris-Buneman integrator (MTS) \cite{Toggweiler2014255}.
MTS integration allows to compute the space charge field less frequently, by a factor which has to be defined beforehand. 
For many situations where space charge has a visible contribution but is not dominant, 
this method can save considerable computation time with only negligible impact on the accuracy. 

For the radio frequency cavities we use a radial voltage profile along the cavity $V(r)$, the gap-width $g$ to correct for the transit time. 
For the time dependent field we get
 
\begin{equation}
 \Delta E_{\RM{rf}} = \frac{\sin\tau}{\tau} \Delta V(r) \cos[\omega_{\RM{rf}} t - \phi],
\end{equation}

with $F$ denoting the transit time factor $F=\frac{1}{2} \omega_{\RM{rf}}  \Delta t$, and
$ \Delta t$ the transit time
\begin{equation}
 \Delta t = \frac{g}{\beta c}.
\end{equation}

In addition, a voltage profile varying along radius will give a phase compression of the bunch, which is induced by an additional magnetic field component $B_{z}$ in the gap,
\begin{equation}
B_{z} \simeq \frac{1}{g \omega_{\RM{rf}} } \frac{d V(r)}{dr} \sin[\omega_{\RM{rf}} t - \phi].
\end{equation}
From this we can calculate a horizontal deflection $\alpha$  as
\begin{equation}
\alpha \simeq \frac{q}{m_{0} \beta \gamma c  \omega_{\RM{rf}}t} \frac{d V(r)}{dr} \sin[\omega_{\RM{rf}} t - \phi].
\end{equation}

In addition, apart from the multi-particle simulation mode, \opalcycl also has two other serial tracking modes for conventional cyclotron machine design. 
One mode is the single particle tracking mode, which is a useful tool for the preliminary design of a new cyclotron. 
It allows to compute  basic parameters, such as reference orbit, phase shift history, stable region and matching phase ellipse.

The other one is the tune calculation mode, which can be used to compute the betatron oscillation frequency $\nu_r$ and $\nu_z$. 
This is useful for evaluating the focusing characteristics of a given magnetic field map.

\subsubsection{Matched distributions}  \label{sssec:matched}
Distributions of beam particles in phase space are matched to
beamlines, if they fit into the acceptance of the beamline. 
In case of circular accelerators like cyclotrons or synchrotons, 
the meaning of a matched distribution does not directly refer to
the acceptance, but to the eigenellipsoid of the accelerator optics.
A distribution is matched to the eigenellipsoid, if it is stationary
on a turn-by-turn basis. In the linear case this can be written as
\begin{equation}
\sigma={\bf M}\,\sigma\,{\bf M}^T\,,
\end{equation}
where $\sigma$ is the matrix of second moments and ${\bf M}$ is
the beam transport matrix for a turn or a sector.
 
In case of cyclotrons the eigenellipsoid is only well-defined for
so-called coasting beams, i.e. for beams that are centered about 
an equilibrium orbit (EO). However, in most cases the change of the 
eigenellipsoid from turn to turn is small and a distribution that
is matched to the eigenellipsoid of the coasting beam is approximately
matched to the accelerated orbit.

In contrast to synchrotons, isochronous cyclotrons have no intrinsic 
longitudinal focusing so that the matching usually concerns the 
transversal coordinates only. This changes in case of high intensity 
beams where the strong space charge forces can induce an effective 
longitudinal focusing in isochronous cyclotrons~\cite{Adam:85,Kos:93}.
While matched eigenellipsoids are - at low beam intensities - properties 
of the accelerator (like the dispersion), the situation becomes more
involved at high intensities, where a realistic transfer map of the 
accelerator depends on the self-forces of the beam. In this case, 
the beam must be matched not only to the accelerator but to the
coupled system of beam and accelerator:
\begin{equation}
\sigma={\bf M}(\sigma)\,\sigma\,{\bf M}(\sigma)^T\,,
\end{equation}
The computation of the transfer maps ${\bf M}$ and matched beam matrices 
$\sigma$ then requires the use of an iterative method even
in the case of a linear approximation~\cite{Bau:2011}.
Since space charge induced self-forces are known to be highly non-linear,
it is not obvious how valuable the results of a linear approximation are.
Therefore the second moments of linearly matched distributions have been 
used to create Gaussian beam distributions, which where used as starting
conditions. 
Besides the proof of principle of the linear matching method, the results
of these simulations yielded further insights into bunch stability conditions, 
namely on the isochronism and the ratio of longitudinal and transverse 
emittance~\cite{Bau:2013}.

\sloppy The linear matching method has been implemented in \opalcycl in order to
provide reasonable starting conditions for matched beam distributions.

\subsubsection{FFA modelling capabilities}  \label{sssec:ffag}

Fixed-Field Alternating Gradient (FFA) machines have been constructed to accelerate electrons and protons over a broad range of energies. FFAs are also well-suited to accelerate unstable particles such as muons due to the potentially rapid acceleration time and applicability in the relativistic regime. Like cyclotrons, FFAs use fixed magnetic fields for focussing. FFAs can have fixed-frequency or variable-frequency RF cavities.

The FFA modelling capabilities of OPAL are implemented as an extension of the
\opalcycl package, OpalRing. Elements can be placed in a 3D world 
volume with placement either as azimuthally connected components or disjoint 
field maps. Field elements can be superimposed, for example in the case where a
FFA magnet overlaps with an RF cavity or neighbouring magnet.

FFA magnets are modelled by a theoretical, perfectly scaling sector field with fringe field, and interpolation from 3D field maps. RF cavities can be modelled using rectangular pillboxes with time varying frequency, phase and amplitude either with or without fringe fields.

For a perfectly scaling sector field, fringe fields off the midplane are calculated to arbitrary order using a recursive power series. Considering a polar coordinate system $(r, \phi, z)$, the field can be written as
\begin{align}
\label{eq:ffa_bz}
B_z &= \sum_{n=0} f_{2n}(\psi) h(r) \left(\frac{z}{r}\right)^{2n} \\
B_\phi &= \sum_{n=0} f_{2n+1}(\psi) h(r) \left(\frac{z}{r}\right)^{2n+1} \\
B_r &= \sum_{n=0}  \left[ \frac{k-2n}{2n+1} f_{2n}(\psi) - \tan(\delta) f_{2n+1} \right] h(r) \left(\frac{z}{r}\right)^{2n+1}
\end{align}
with $\psi$ representing the azimuthal angle in spiral coordinates
\begin{equation}
\label{eq:ffa_angle}
\psi = \phi - \tan(\delta) \ln(r/r_0),
\end{equation}
 and $h(r)$ the field dependence on radius
 \begin{equation}
h(r) = B_0 \left(\frac{r}{r_0}\right)^k.
\end{equation}
The fringe field on the midplane is modelled using
\begin{equation}
f_0(\psi)=\frac{1}{2}\left[\tanh \left(\frac{\psi+\psi_0}{\lambda}\right)-\tanh \left(\frac{\psi-\psi_0}{\lambda} \right)\right]
\end{equation}
so that $\psi_0$ represents the length of the flat-top and $\lambda$ represents the length of the field fall-off.  Away from the midplane, the coefficients $f_{n}$ are calculated by requiring that Maxwell's equations are observed, yielding a recursion relation in terms of the derivatives of $f_{0}$
\begin{align}
f_{n} &= \sum_{i=0} a_{i,n} \partial^i_\psi f_{0} \\
\end{align}
with even and odd terms related by
\begin{align}
a_{i,2n+1} =& \frac{a_{i-1,2n}}{2n+1} \\
a_{i,2n+2} =& \frac{1}{2n+2} \big( a_{i,2n+1} 2(k-2n) \tan(\delta)  \\
            &- \frac{(k-2n)^2}{2n+1} a_{i, 2n} - (1+\tan^2(\delta)) a_{i-1, 2n+1} \big).
\end{align}

$f_0(\psi)$ is implemented using C++ inheritance making alternate fringe field models, for example Enge function, simple to implement. The coefficients $f_n$ are calculated during the lattice construction to minimise overhead during stepping.

FFA fields can also be simulated using 3D field maps generated by external field modelling tools. A trilinear interpolation is applied to get field maps between grid points.

RF cavities are modelled using a sinusoidally varying pill box model, with fields given by
\begin{equation}
\vec{E} = E_0 \sin(\omega t + \phi) \hat{z}
\end{equation}
Hard-edged cavities are assumed to have constant electric fields across the cavity aperture and the magnetic fields are assumed to be negligible. Frequency $\omega$, phase $\phi$ and peak field $E_0$ can be varied by supplying a fourth order polynomial in time, enabling tuning of the accelerating frequency and bucket height to match the variable time-of-flight characteristic of FFAs.

Soft-edged cavities are described in Cartesian coordinates $(x, z, s, t)$ by a series expansion off the midplane,
\begin{align}
E_x & = 0 \\
E_z & = V(t) \sum_n z^n g_n \sin(\Omega t + \Phi) \\
E_s & = V(t) \sum_n z^n f_n \sin(\Omega t + \Phi) \\
B_x & = V(t) \sum_n y^n h_n \cos(\Omega t + \Phi) \\
B_y & = 0 \\
B_z & = 0.
\end{align}
It is assumed that variations in $\Omega$, $\Phi$ and $V$ occur on a timescale much longer than the RF frequency $\Omega$ so that they do not contribute significantly to the magnetic field. Recursion relations can be derived assuming Maxwell's laws,
\begin{align}
g_{n+1} &= -\frac{1}{n+1}\partial_z f_n \\
h_{n+1} &= -\frac{\omega}{c^2 (n+1)} f_n \\
f_{n+2} &= -\frac{\omega^2 f_n +  c^2 \partial^2_z f_n}{c^2(n+1)(n+2)}.
\end{align}
Odd terms of $f_n$ and even terms of $g_n$ and $h_n$ are 0.

\subsection{Space charge models }  \label{ssec:sc}
\subsubsection{FFT Based solver}  \label{sssec:fft}

The Particle-Mesh (PM) solver is one of the oldest improvements over the PP solver. Still one of the best references is the book by R.W.~Hockney \& J.W.~Eastwood \cite{hockney}.
The PM solver introduces a discretisation of space. The rectangular computation domain $\Omega:=[-L_x,L_x]\times[-L_y,L_y]\times[-L_t,L_t]$, just big enough to include all particles, is segmented into a regular mesh of $M=M_x\times M_y\times M_t$ grid points. For the discussion below we assume $N=M_x=M_y=M_t$.

The solution of Poisson's equation is an essential component of any self-consistent electrostatic beam dynamics code that models the transport of intense charged particle beams in accelerators. If the bunch is small compared to the transverse size of the beam pipe, the conducting walls are usually neglected.
In such cases the Hockney method may be employed \cite{hockney, eastwoodandbrownrigg,hockneyandeastwood}. In that method, rather than computing $N_p^2$ point-to-point interactions
(where $N_p$ is the number of macroparticles), the potential is instead calculated on a grid of size $(2 N)^d$, where $N$ is the number of grid points in each dimension of the physical mesh containing
the charge, and where $d$ is the dimension of the problem.
Using the Hockney method, the calculation is performed using Fast Fourier Transform (FFT) techniques, with the computational effort scaling as $(2N)^d (log_2 2N)^d$.

When the beam bunch fills a substantial portion of the beam pipe transversely, or when the bunch length is long compared with the pipe transverse size, the conducting boundaries cannot be ignored. Poisson solvers have been developed previously to treat a bunch of charge in an open-ended pipe with various geometries \cite{qiangandryne,qiangandgluckstern}.

The solution of the Poisson equation,

\begin{equation}
\nabla^2\phi=-\rho/\epsilon_0,
\end{equation}
for the scalar potential, $\phi$, due to a charge density, $\rho$, and appropriate boundary conditions, can be expressed as,

\begin{equation}
\phi(x,y,z)=\int\int\int{dx' dy' dz'}\rho(x',y',z') G(x,x',y,y',z,z'),
\end{equation}
where $G(x,x',y,y',z,z')$ is the Green function, subject to the appropriate boundary conditions, describing the contribution of a source charge at location $(x',y',z')$ to the potential at an observation location $(x,y,z)$.

For an isolated distribution of charge this reduces to

\begin{equation}
\phi(x,y,z)=\int\int\int{dx' dy' dz'}\rho(x',y',z') G(x-x',y-y',z-z'),
\label{convolutionsolution}
\end{equation}
where

\begin{equation}
G(u,v,w)={\frac{1}{\sqrt{u^2+v^2+w^2}}}.
\label{isolatedgreenfunction}
\end{equation}
A simple discretisation of Eq.~(\ref{convolutionsolution})
on a Cartesian grid with cell size $(h_x,h_y,h_z)$
leads to,

\begin{equation}
\phi_{i,j,k}=h_x h_y h_z \sum_{i'=1}^{M_x}\sum_{j'=1}^{M_y}\sum_{k'=1}^{M_t}  \rho_{i',j',k'}G_{i-i',j-j',k-k'},
\label{openbruteforceconvolution}
\end{equation}
where $\rho_{i,j,k}$ and $G_{i-i',j-j',k-k'}$ denote the values of the charge density and the Green function, respectively, defined on the grid $M$.

\paragraph{FFT-based Convolutions and Zero Padding}
FFTs can be used to compute convolutions by appropriate zero-padding of the sequences.
Discrete convolutions arise in solving the Poisson equation, and one is typically interested in the following,

\begin{equation}
\bar{\phi}_j=\sum_{k=0}^{K-1}\bar{\rho}_k \bar{G}_{j-k}\quad,
\begin{array}{l}
j=0,\ldots,J-1 \\
k=0,\ldots,K-1 \\
j-k=-(K-1),\ldots,J-1 \\
\end{array}
\label{bruteforceconvolution}
\end{equation}
where $\bar{G}$ corresponds to the free space Green function, $\bar{\rho}$ corresponds to the charge density, and $\bar{\phi}$ corresponds to the scalar potential.
The sequence $\{\bar{\phi}_j\}$ has $J$ elements, $\{\bar{\rho}_k\}$ has $K$ elements, and $\{\bar{G}_m\}$ has $M=J+K-1$ elements.

One can zero-pad the sequences to a length $N\ge M$ and use FFTs to efficiently obtain the $\{\bar{\phi}_j\}$ in the unpadded region.
This defines a zero-padded charge density, $\rho$,
\begin{equation}
\rho_k=\left\{
\begin{array}{l l}
\bar{\rho}_k & \quad \text{if }k=0,\ldots,K-1 \\
0 & \quad \text{if }k=K,\ldots,N-1. \\
\end{array}\right.
\end{equation}
Define a periodic Green function, $G_m$, as follows,
\begin{equation}
G_m=\left\{
\begin{array}{l l}
\bar{G}_m & \quad \text{if }m=-(K-1),\ldots,J-1 \\
0 & \quad \text{if }m=J,\ldots,N-K, \\
G_{m+iN}=G_{m} & \quad \text{for } i \text{ integer }.
\end{array}\right.
\label{periodicgreenfunction}G
\end{equation}
Now consider the sum
\begin{equation}
{\phi}_j={1\over N}\sum_{k=0}^{N-1} W^{-jk}
                    \left(\sum_{n=0}^{N-1} \rho_n W^{nk}\right)
                    \left(\sum_{m=0}^{N-1} G_m W^{mk}\right),
~~~~~~0 \le j \le N-1,
\label{fftconvolution}
\end{equation}
where $W=e^{-2\pi i/N}$. This is just the FFT-based convolution of $\{\rho_k\}$ with $\{G_m\}$.
Then,
\begin{equation}
{\phi}_j=
          \sum_{n=0}^{K-1}~
          \sum_{m=0}^{N-1} \bar{\rho}_n G_m
{1\over N}\sum_{k=0}^{N-1} W^{(m+n-j)k}
~~~~~~0 \le j \le N-1.
\end{equation}
Now use the relation
\begin{equation}
\sum_{k=0}^{N-1} W^{(m+n-j)k}= N \delta_{m+n-j,iN}~~~~~(i~\rm an~integer).
\end{equation}
It follows that
\begin{equation}
{\phi}_j=\sum_{n=0}^{K-1}~\bar{\rho}_n G_{j-n+iN}
~~~~~~0 \le j \le N-1.
\end{equation}
But $G$ is periodic with period $N$. Hence,
\begin{equation}
{\phi}_j=\sum_{n=0}^{K-1}~\bar{\rho}_n G_{j-n}
~~~~~~0 \le j \le N-1.
\label{finaleqn}
\end{equation}
In the physical (unpadded) region, $j\in \left[0,J-1\right]$, so the quantity $j-n$ in Eq.~(\ref{finaleqn}) satisfies $-(K-1)\le j-n \le J-1$.
In other words the values of $G_{j-n}$ are identical to $\bar{G}_{j-n}$. Hence, in the physical region the FFT-based convolution, Eq.~(\ref{fftconvolution}),
matches the convolution in Eq.~(\ref{bruteforceconvolution}).

As stated above, the zero-padded sequences need to
have a length $N \ge M$, where $M$ is the number of elements in the Green function sequence $\left\{x_m\right\}$.
In particular, one can choose $N=M$, in which case the Green function sequence is not padded at all, and only
the charge density sequence, $\left\{r_k\right\}$, is zero-padded, with $k=0,\ldots,K-1$ corresponding to the physical region
and $k=K,\ldots,M-1$ corresponding to the zero-padded region.


The  above FFT-based approach -- zero-padding the charge density array, and circular-shifting the Green function in accordance with Eq.~(\ref{periodicgreenfunction}) -- will work in general.
In addition, if the Green function is a symmetric function of its arguments,  the value at the end of the Green function array
(at grid point $J-1$)
can be dropped, since it will be recovered implicitly through the symmetry of Eq.~(\ref{periodicgreenfunction}).
In that case the approach is identical to the Hockney method \cite{hockney, eastwoodandbrownrigg,hockneyandeastwood}.

Lastly, note that the above proof that the convolution,  Eq.~(\ref{fftconvolution}), is identical to Eq.~(\ref{bruteforceconvolution}) in the unpadded region, works even when
$W^{-j k}$ and $W^{m k}$  are replaced by $W^{j k}$ and $W^{-m k}$, respectively, in Eq.~(\ref{fftconvolution}). In other words, the FFT-based approach can be used to compute
\begin{equation}
\bar{\phi}_j=\sum_{k=0}^{K-1}\bar{\rho}_k \bar{G}_{j+k}\quad,
\begin{array}{l}
j=0,\ldots,J-1 \\
k=0,\ldots,K-1 \\
j-k=-(K-1),\ldots,J-1 \\
\end{array}
\label{bruteforcecorrelation}
\end{equation}
simply by changing the direction of the Fourier transform of the Green function and changing the direction of the final Fourier transform.

\paragraph{Algorithm used in \opal}

As a result, the solution of Eq.~(\ref{openbruteforceconvolution}) is then given by

\begin{equation}
\phi_{i,j,k}=h_x h_y h_z \text{FFT}^{-1} \{ ( \text{FFT}\{\rho_{i,j,k}\}) ( \text{FFT}\{G_{i,j,k}\}) \}
\label{oneterm}
\end{equation}
where the notation has been introduced that $\text{FFT}\{ . \}$ denotes a forward FFT in all 3 dimensions,
and $\text{FFT}^{-1}\{ . \}$  denotes a backward FFT in all 3 dimensions.

\paragraph{Interpolation Schemes}
\label{sec:interpol}
Both charge assignment and electric field interpolation are related to the interpolation
scheme used. A detailed discussion is given in~\cite{hockney}.
If $e_i$ is the charge of a particle, we can write the density at mesh point $\vec{k}_m$ as
\begin{equation}\label{eq:discRho}
\rho(\vec{k}_m)^D = \sum_{i=1}^N e_i\cdot W(\vec{q}_i,\vec{k}_m), ~ m=1\dots M
\end{equation}
where $W$ is a suitably chosen weighting function (with local support).
The simplest scheme is the nearest grid point (NGP) method, where the total particle charge is assigned to
the nearest grid point and the electric field is also evaluated at the nearest grid point. A more
elaborate scheme is called cloud in cell (CIC). It assigns the charge to the $2^d$ nearest grid points and
also interpolates the electric field from these grid points. The assigned density changes are continuous when
a particle moves across a cell boundary, although the first derivative is discontinuous. 

\subsubsection{SAAMG solver}  \label{sssec:saamg}

\begin{figure}[htb]
  \centering
  \includegraphics[width=0.3\textwidth]{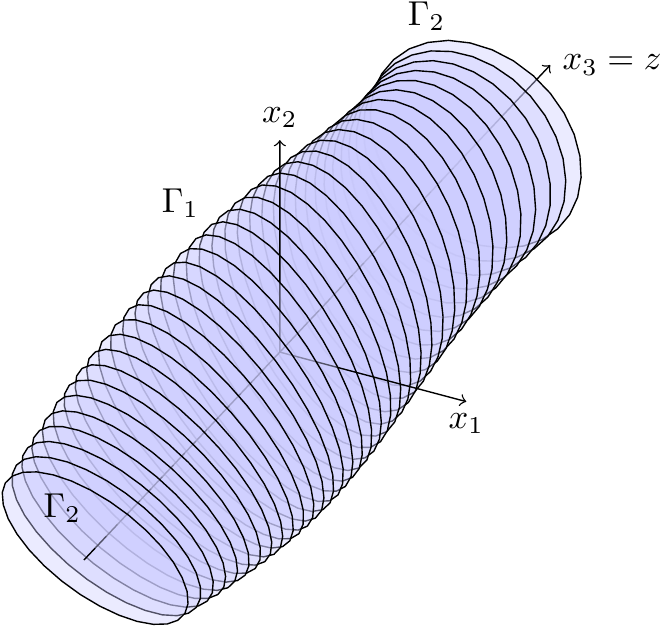}
  \caption{Sketch of a typical domain}
  \label{fig:domain}
\end{figure}

As we have previously seen the PIC approach allow us to solve the space charge
  Poisson problem.
Unfortunately some advantageous properties of the slowly varying fields
  and repetitive calls to the solver cannot be exploited by PIC approaches.
In addition, this approach is limited to simple geometries such as boxes and
  cylinders.
In order to take into account the non-linear beam dynamics, which is
  indispensable for the design of next generation of particle accelerators,
  we require better models for the image charges on the true geometry of the
  beam-pipe as visualized in Figure~\ref{fig:domain}.

In the rest of this section we describe how the following Poisson equation
\begin{equation} \label{eq:poisson}
  \begin{aligned}
    -\Delta \phi &= \frac{\rho}{\epsilon_0}\ \text{in}\ \Omega, \\
    \phi &= g \equiv 0\ \text{on}\ \Gamma_1,   \\
    \frac{\partial \phi}{\partial \mathbf{n}} + \frac {1}{d} \phi &= 0\
    \text{on}\ \Gamma_2
  \end{aligned}
\end{equation}
  is solved based on~\cite{Adelmann20104554} and its
  improvements~\cite{para2010}.

\paragraph{The Discretization}

We discretize~\eqref{eq:poisson} by a second order finite difference
  scheme defined on a rectangular lattice (grid) with a regular 7-point
  difference stencil for interior points.
Near the boundary we have to take the boundary conditions into account.
The beam pipe boundary is modeled using Dirichlet boundary conditions
  ($\Gamma_1$), and to model the open ends of the beam pipe (longitudinal
  direction) we use Robin boundary conditions ($\Gamma_2$).
Generally the intersection of the beam pipe and our lattice does not collide
  with grid points.
To handle these cases we implemented different extrapolation schemes, where we assign the one of the
  following values to the next grid point on the outside:
\begin{itemize}
  \item Constant extrapolation: use the boundary value.
  \item Linear extrapolation: linearly extrapolate between the last interior
        point and the boundary value.
  \item Quadratic (Shortley-Weller) extrapolation: same as linear, quadratic
        using the last two interior points and the boundary value.
\end{itemize}

Together with the discretization matrix $A$ we can formulate the system
  $A \mathbf{x} = \mathbf{b}$ that can be solved with any iterative approach.
Here $\mathbf{x}$ is the vector of unknown values of the potential and
  $\mathbf{b}$ is the vector of the charge density interpolated at the grid
  points.
Notice that the quadratic extrapolation at the boundary yields a
  \emph{nonsymmetric} but still positive definite Poisson matrix.

\paragraph{The Solution Method}

We observed good convergence and scaling of the solver when employing a
  preconditioned Conjugate Gradient method.
The preconditioned system of the discretized problem has the form
\begin{equation*}
  {M}^{-1}{A} \mathbf{x} = {M}^{-1}\mathbf{b},
\end{equation*}
where the positive definite matrix $M$ is the preconditioner.
Good preconditioners reduce the condition number of the system and thus the
  number of steps the iterative solver takes until
  convergence~\cite{hack:94,gree:97}.
Employing a preconditioner is inevitable for systems originating in finite
  difference discretizations of 2nd order PDE's, that we have to handle here,
  since their condition number increases as $h^{-2}$ where $h$ is the mesh
  width~\cite{leve:07}.

In our solver we opted to use a smoothed aggregation-based algebraic Multigrid
  (AMG) preconditioner.
In general Multigrid or multilevel preconditioners are the most effective
  preconditioners, in particular for the Poisson problem~\cite{hack:85,tros:00}.
With smoothed aggregation-based AMG preconditioning the aggregation smoother
  takes care of anisotropies and related issues and leads to a robustness
  superior to geometric multigrid, see~\cite{trcl:09} for a discussion.
The preconditioner easily adapts to the elongation  of the computational domain
  that happens during our simulation.

Both, the preconditioner and the iterative solver are based on the Trilinos
  framework~\cite{Trilinos-Web-Site, Trilinos-TOMS}.
We tuning the involved algorithmic parameters of multigrids, such as for
  example the coarse level solver, restriction and prolongation operators is
  key in attaining performance.

Since we have to solve a sequence of Poisson problems, and space charge forces
  generally do not change radically from one time step to the next, we showed
  that the performance (time to solution) can be improved by
\begin{itemize}
  \item using the solution from the last time step as a starting point, and
  \item reusing the hierarchy of the multigrid of the previous time step, or
  \item reusing the preconditioner of the previous time step.
\end{itemize}
In order to avoid degeneration, we recompute the hierarchy or preconditioner
  once the total number of CG iterations has increased by a certain threshold.

In ~\cite{Adelmann20104554} we show in a real world example the relevance of this approach,
by observing up to
  $~40\%$ difference in the rms beam size when comparing to the FFT-based
  solver with open domains.
With the modifications reported in~\cite{para2010} the solver scaled well up
  to 2048 processors ($83\%$ parallel efficiency starting from 512) with
  cylindrical tubes embedded in meshes with up to $1024^3$ grid points.

\subsubsection{AMR solver}  \label{sssec:amr}
The usage of regular grids for problems that are very heterogeneous with
respect to the spatial distribution of particles isn't advantageous.\
Only small areas of interest require a fine resolution while much coarser
grids are sufficient to resolve the larger parts of a problem.\ This
observation leads to the technique of adaptive mesh refinement (AMR).\

In order to benefit from this numerical technique we enhanced the
mesh and particle handling in \opal with the memory efficient and
scalable \amrex~\cite{AMReX} (successor of \boxlib~\cite{BoxLib})
framework.\ The flexible grid adaptation is enabled only in the
multi-bunch mode of \opalcycl where it takes into account either
the field strength (i.e.\ scalar potential or electric field) or
the charge density.\

In opposite to all other space charge solvers in \opal the mesh size
of each level of refinement stays constant during the evolution of
the macro particles.\ Instead, in every time step the particles are
mapped into a predefined domain with default $[-1, 1]^3$ where the Poisson problem~\eqref{eq:poisson}
is solved with either periodic, Dirichlet or open boundary conditions.\
The fields are later properly rescaled to obtain the solution of the
original problem.\

Beside a built-in AMR framework solver the new feature provides a multigrid
solver \cite{2018arXiv181203689F} that is based on second
generation packages of \trilinos{} \cite{1089021}.

\subsection{Particle matter interaction model}  \label{ssec:patriot}
One of the unique features of \opal\ is to perform not only the particle tracking through an accelerator or beam line, but also a Monte Carlo simulation of the beam interaction with matter.
A fast charged particle incident on a slab of material undergoes collisions with the atomic electrons losing energy and with nuclei being scattered away from its original trajectory. Both of these processes are implemented in \opal, namely the energy loss with the Bethe-Bloch formula and the path deflection with the Multiple Coulomb Scattering (MCS) and Single Rutherford Scattering (SRS). Performing the calculations parallel on multiple cores, \opal\ tracks particles through the accelerator and integrates the equations of motion in a specified time step $\Delta t$, along the particle trajectory. In addition, the process  \texttt{surface physics} \cite{opal}  applied to an element of the lattice (such as a collimator) allows the users to include into the tracking a Monte Carlo simulation of the particles interaction with that element.  It has to be remarked that, at the moment, all the physical processes in \opal\  are restricted only to the proton interaction with matter.

\subsubsection{Energy loss}
The loss of energy by the incident particle occurs almost entirely in collisions with the electrons in the medium. Following the ICRU 49 guideline \cite{ICRU}, the stopping power is defined by two different models according to the initial energy (or momentum) of the incident particle.  At low energies, the electronic stopping powers are obtained from experimental data and empirical formulas developed by Andersen and Ziegler for protons. The border between the high- and low- energy region depends on the accuracy of the corrections to the Bethe-Bloch formula available for various materials. In case of protons, this border has been fixed around to 0.6 MeV. The models for high- and low- energies are implemented in \opal\ as well, in order to cover a large spectrum of possible scenarios. In the theory of Andersen and Ziegler, the independent variable is the scaled energy $T_s$ which is equal to the kinetic energy T (in keV) divided by the ratio of the proton mass to the atomic mass unit ($M_p/u$ = 1.0073) \cite{ICRU}. The stopping cross section ($T_s$) is fitted by the equation: 
\begin{equation}
\epsilon = \frac{\epsilon_{low}\cdot \epsilon_{high}}{\epsilon_{low}+\epsilon_{high}}
\label{eq:epsilon}
\end{equation} 
where 
\begin{equation}
\epsilon_{low} = A_2T_s^{0.45}
\label{eq:eps_low}
\end{equation}
and 
\begin{equation}
\epsilon_{high} = \frac{A_3}{T_s}\ln{\left(1+\frac{A_4}{T_s} + A_5T_s\right)}
\label{eq:eps_high}
\end{equation}
The numerical values of the coefficients $A_i$ (i = 2...5) are specific of each material and adjusted with experimental data. The corresponding stopping power for low energy protons is given by
\begin{equation}
-\left\langle \frac{dE}{dx} \right\rangle = -\frac{\epsilon}{(A/N_A)\cdot10^{21}}
\label{eq:Bethe_low}
\end{equation}
where A is the atomic mass of the medium and $N_A$ the Avogadro's number. For each material included in \opal, the corresponding coefficients $A_i$ are implemented as well and the stopping power for low energy protons is then evaluated with \qqref{eq:Bethe_low}. For energy higher that 0.6 MeV, the energy loss is calculated by \opal\ using the following Bethe-Bloch equation
\begin{equation}
-\left\langle \frac{dE}{dx} \right\rangle = Kz^2 \frac{Z}{A}\frac{1}{\beta^2}\left[\frac{1}{2}ln\left(\frac{2m_ec^2\beta^2\gamma^2W_{max}}{I^2}\right)-\beta^2\right]
\label{eq:Bethe}
\end{equation}
where z is the charge of the incident particle and $Z, A$ are respectively the atomic number, atomic mass of the absorber. For the mean excitation energy, $I$, \opal\ uses the equations in \cite{Leo}, where $I$ depends on the atomic number of the absorber, such that:
\begin{eqnarray}
\frac{I}{Z} &=& 12 + \frac{7}{Z} , \quad \text{for }   Z < 13 \\
\frac{I}{Z} &=& 9.76 + 58.8\cdot Z^{-1.19}  \quad \text{for } Z \geq 13
\end{eqnarray}
$\beta$ and $\gamma$ are kinematic variables and K is a constant given by: 
\begin{equation}
K = 4\pi N_Ar_e^2m_ec^2
\end{equation}
where $r_e$ is the classical electron radius and $m_e$ the electron mass. \\
In \qqref{eq:Bethe}, $W_{max}$ is the maximum kinetic energy which can be imparted to a free electron in a single collision and is expressed by:
\begin{equation}
W_{max} = \frac{2m_ec^2\beta^2\gamma^2}{1+2\gamma m_e/M +(m_e/M)^2}
\end{equation}
where M is the mass of the incident particle. \\
The Bethe-Bloch equation implemented in \opal (\qqref{eq:Bethe}) does not contain the density-effect correction $\delta$. This factor describes the reduction of the stopping power due to the polarisation of the medium and it is important only in relativistic regime. Therefore, for proton therapy application, neglecting this term is perfectly reasonable. For relatively thick absorbers such that the number of collisions is large, the energy loss distribution has a Gaussian shape: 
\begin{equation}
f(x,\Delta) \propto  e^{-\frac{(\Delta - \bar{\Delta})^2}{2\sigma}}
\end{equation}
where $\Delta$ is the energy loss in the absorber, $\bar{\Delta}$ the mean energy loss and x the absorber thickness. For non-relativistic heavy particles, the $\sigma$ of the Gaussian distribution is calculated by
\begin{equation}
\sigma^2 = 4\pi N_a r_e^2(m_ec^2)^2\rho\frac{Z}{A}x
\label{eq:sigma}
\end{equation}
where $\rho$ is the target density.
In \opal\ algorithm, when the surface physics card is selected and a particle hits a material, the energy loss is calculated using \qqref{eq:Bethe_low} or (\ref{eq:Bethe}) \cite{Stachel}. So a particle, travelling through a material of density $\rho$ in a single step $\Delta s$, looses an average energy given by

\begin{equation}
-\left\langle \frac{dE}{dx} \right\rangle \cdot \Delta s \cdot \rho = \overline{dE}
\end{equation}

The actual energy loss is calculated adding to $\overline{dE}$ a value obtained from a Gaussian random distribution of mean $\overline{dE}$ and width $\sigma_E$ (\qqref{eq:sigma}) for a target thickness equal to $\Delta s$. In addition, \opal removes a particle from the bunch if its energy is less than 0.1 MeV.

\subsubsection{Coulomb scattering}
A charged particle traversing a medium is deflected from its original trajectory due to the Coulomb scattering with nuclei. Depending on the angular deflection, the Coulomb scattering is referred as: 

\begin{itemize}
\item \textbf{Multiple Coulomb Scattering} in case of many small angles. The probability of this process is quite large and the net result is Gaussian angular distribution. The MCS is well described by Moliere's theory. 
\item \textbf{Single Rutherford Scattering} in case of a single large angle. The probability of this process is low and the net results are tails in the angular distribution. 
\end{itemize}

Following \cite{Jackson}, it is more convenient to express the scattering angle in terms of the relative projected angle

\begin{equation}
 \alpha = \frac{\theta }{<\Theta ^2> ^{1/2}} 
\end{equation}

where $\theta $ is the angle of scattering and $<\Theta ^2>$ is the mean square angle. The multiple- and single- scattering distributions can be written as

\begin{equation}
 P_{Multi}(\alpha) d\alpha = \frac{1}{\sqrt{\pi}}e^{-\alpha^2} d\alpha
\label{eq:PM}
\end{equation}

\begin{equation}
 P_{Single}(\alpha) d\alpha = \frac{1}{8\cdot \ln{(204Z^{-1/3})}}\frac{d\alpha}{\alpha^3}
\label{eq:PS}
\end{equation}

Combining \qqref{eq:PM} and (\ref{eq:PS}), the complete angular distribution shows a Gaussian core with lateral tails due to the single scattering. 

The transition from multiple- to single- scattering occurs for

\begin{equation}
\alpha = \frac{\theta}{<\Theta ^2> ^{1/2}} = \frac{\theta}{\sqrt{2}\cdot \theta_0} = 2.5 \rightarrow \theta = 3.5 \theta_0
\label{eq:alpha_thre}
\end{equation}

 where $\theta_0$ is the scattering angle according to the Moliere's theory and expressed by

\begin{equation}
 \theta_0 = \frac{13.6 MeV}{\beta c p}z\sqrt{\frac{\Delta s}{X_0}}\left[1+0.038\ln{\left(\frac{\Delta s}{X_0}\right)}\right]
\end{equation}

where $X_0$ is the radiation length of the target material and $\Delta s$ is the step width in space.  

\subsubsection{Multiple Coulomb Scattering}

The model in \opal for the MCS follows the description of the \textit{Particle Data Group} \cite{PDG}. 



The main parameters for evaluating the effect of the MCS are the new spatial coordinate $y_{plane}$ and the new angle $\theta_{plane}$, defined by

\begin{equation}
y_{plane} = \frac{z_1 \theta_0 \Delta s}{\sqrt{12}} + \frac{z_2\theta_0 \Delta s}{2}
\end{equation}

and 

\begin{equation}
\theta_{plane} = z_2 \theta_0
\end{equation}

where $z_1$ and $z_2$ are independent numbers from a Gaussian random distribution with mean zero and variance one. 

Once the MCS angle has been evaluated, the reference system of the particle is adjusted in \opal to the new direction of motion with the angle $\Psi_{yz}$ (see the Master Thesis of H. Stachel for the full calculation~\cite{Stachel}). The new coordinates of the particle are

\begin{equation}
y = y +\Delta s \cdot p_y + y_{plane}\cdot \cos{\Psi_{yz}}
\end{equation}

\begin{equation}
z = z -y_{plane}\cdot \sin{\Psi_{yz}} + \Delta s \cdot p_z
\end{equation} 

The new coordinates in the x-plane are analogous. 

\subsubsection{Single Rutherford Scattering}

The percentage of particles undergoing large angle scattering is

\begin{equation}
\chi_{single} = \frac{\int_{2.5}^{\infty}P_{Single}(\alpha)d\alpha}{\int_{0}^{2.5}P_{Multiple}(\alpha)d\alpha+\int_{2.5}^{\infty}P_{Single}(\alpha)d\alpha}
\end{equation}

where $P_{Multiple}$ and $P_{Single}$ are given by \qqref{eq:PM} and (\ref{eq:PS}) respectively. In general, $\chi_{single}$ does not change that much with the atomic number Z of the target. Its value is around 0.004 - 0.005 for many materials. In \opal algorithm, a fixed value of 0.0047 is used for $\chi_{single}$.  A random number $\xi_1$ between 0 and 1 is generated and if it is smaller than 0.0047, the particle undergoes single Rutherford scattering. \\

Rutherford angle is calculated with a second random number $\xi_2$ between 0 and 1 for deciding how larger is the angle with respect to $2.5 \alpha$ (or $3.5 \theta_0$ according to the critical value of \qqref{eq:alpha_thre}). 

\begin{equation}
\theta_{Rutherford} = \pm 2.5\cdot \sqrt{\frac{1}{\xi_2}}\cdot \theta_0
\end{equation}

where the sign is given by a third random number needed to determine the scattering direction (up- or downwards).

\subsection{Beam residual interactions}  \label{ssec:resgas}
The study of beam losses plays an important role in accelerator design with regards to optimizing transmission as well as minimizing activation associated with lost particles. OPAL has integrated residual physics phenomena important to this problem. The processes with the most relevance are the interaction of the beam with residual gas and electromagnetic beam charge stripping. Since these phenomena are stochastic in nature, a Monte Carlo method is employed. Also, the focus is on $H^-$ and $H^+$ beams, as electromagnetic stripping is used extensively with these types of particles. However, the model can be extended to other ions, such as $H_2^+$.

Assuming that particles are normally incident on a homogeneous medium and that they are subject to a process with a mean free path $\lambda$ between interactions, the probability density function for the interaction of a particle after travelling a distance $x$ is \cite{Leo,Tavernier}:
\begin{equation}
F(x)=\frac{1}{\lambda}\cdot\,e^{\,-x/\lambda}
\label{eq:pdf}
\end{equation}
where $F\!(x)dx$ represents the probability of having an interaction between $x$ and $x\!+\!dx$. From this, the probability of suffering an interaction before reaching a path length $x$ can be deduced:
\begin{equation}
P(x) \,= \displaystyle\int_{0}^{x} F(x)dx \;=\; 1 - e^{-x/\lambda}
\label{eq:int_prob}
\end{equation}
where $P(x)$ is the cumulative interaction probability of the process. In the case of interaction between a beam with particles in a material, the process is generally described in terms of the cross section, $\sigma$. In this case, the mean free path is related to the density of interaction centers and the cross section: $\lambda=1/N\sigma$.

The beam stripping model implemented in OPAL considers interactions for hydrogen ions and protons. In the case of hydrogen, the processes are classified according to the final charge state of the hydrogen, $\sigma_{qq'}$, where $qq'$ represents any one of the combinations for charge state. Concretely, $H^-$ ions, one of the most accelerated particles in cyclotrons, have two electrons: one tightly bound with a binding energy of $13.59843449\,(8)$ eV \cite{Mohr} and another one slightly bound with a binding energy of $0.75419\,(2)$ eV \cite{lykke}. The processes to be considered in this case are single- or double-electron-detachment ($\sigma_{-10}$ or  $\sigma_{-11}$). In the case of protons, the process available is electron capture.

\subsubsection{Residual gas interaction}
Assuming a beam flux incident in an ideal gas with density $N$ (number of gas molecules per unit volume), the number of particles interacting depends on the gas composition and the different reactions to be considered. Using Dalton's law of partial pressures:
\begin{equation}
\frac{1}{\lambda_{total}}\!=\!\!\sum\! \frac{1}{\lambda_k}\!=\!N_{total}\cdot\sigma_{total}=\!\!\sum_j\!N_j\,\sigma^{,j}_{total}\!=\!\!\sum_j\!\left(\!\sum_iN_j\,\sigma^{,j}_{i}\right)
\end{equation}
where the first summation is over all gas components and the second summation is over all processes for each component. Thus, the fraction loss of the beam for unit of travelled length will be, according to \qqref{eq:int_prob}:
\begin{equation}
f_g=1-e^{-x/\lambda_{total}}
\end{equation}

The cross sections for considered ions have been measured experimentally for the most important gases \cite{Whittier,Allison,Hvelplund,Huq,nakai1,nakai2,phelpsH2,phelpsAr,McClure,Barnet,Williams,berkner}. In addition, analytic expressions fitted to cross section data have been semi-empirically developed \cite{Green} from functional forms for collisions of hydrogen atoms and ions with some gases \cite{Nakai}:
\begin{equation}
\sigma_{qq'} = \sigma_0 \left[ f(E_1) + a_7\!\cdot\!f(E_1/a_8) \right]
\label{eq:sigma_nakai}
\end{equation}
where $\sigma_0$ is a convenient cross section unit ($\sigma_0 = 1\cdot 10^{-16}\;\text{cm}^2$); and $f(E)$ and $E_1$ are given by:
\begin{equation}
f(E) = \frac{ a_1\!\cdot\!\left(\!\displaystyle\frac{E}{E_R}\!\right)^{\!a_2} }{ 1+\left(\!\displaystyle\frac{E}{a_3}\!\right)^{\!a_2+a_4}\!\!+\left(\!\displaystyle\frac{E}{a_5}\!\right)^{\!a_2+a_6} }
\end{equation}
\begin{equation}
E_R = hcR_{\infty}\!\cdot\!\frac{m_H}{m_e} = \frac{m_He^4}{8\varepsilon_0^2h^2}
\end{equation}
\begin{equation}
E_1 = E_0 - E_{th}
\end{equation}
where $E_0$ is the incident projectile energy in keV, $E_{th}$ is the threshold energy of reaction in keV, and the symbols $a_i$ ($i\!=\!1,...,8$) denote adjustable parameters.
The OPAL algorithm includes cross sections resulting from \qqref{eq:sigma_nakai}, although this function has been improved, in the case of reactions with $H_2$, to obtain a better fit with experimental data \cite{TabataShi}. This model can be extended to include other types of residual gases.

\subsubsection{Electromagnetic stripping}
When a charged particle (other than an electron or a proton) is in a magnetic field, the electron(s) and the nucleus are bent in opposite directions according to their electric charge. In the case where there is a slightly bound electron, such as $H^-$, a strong enough magnetic field can strip the electron. The lifetime of such particles in an electromagnetic field are of crucial interest in accelerators, specially in high magnetic field and high energy accelerators. (In OPAL, electromagnetic stripping is only implemented for $H^-$ beams.)

The magnetic field in an accelerator, concretely the orthogonal component of the field to the velocity of particles, produces an electric field in the rest frame of the beam according to the Lorentz transformation, $E\!=\!\gamma\beta cB$. Electromagnetic stripping for ions in an accelerator an be analysed as the decay of an atomic system in a weak and static electric field.

The fraction of particles dissociated by the electromagnetic field after a distance travelled $x$ during a time $t$ is a function of energy and magnetic field intensity $B$, according to \qqref{eq:int_prob}:
\begin{equation}
f_{em}=1-e^{\,-\,x/\beta c\gamma\tau}=1-e^{\,-\,t/\gamma\tau}
\end{equation}
where $\tau$ is the particle lifetime in the rest frame. Thus, particle lifetime in the laboratory frame, $\tau_0$ , is related to particle lifetime in the rest frame $\tau$ by Lorentz transformation: $\tau_0\!=\!\gamma\!\cdot\!\tau$. Hence, the mean free path, $\lambda$, in the laboratory frame is given by the relation $\lambda=\beta c\gamma\tau=\beta c\tau_0$.

Theoretical studies \cite{Scherk} have determined the lifetime $\tau$ of an $H^-$ ion in an electric field $E$:
\begin{equation}
\tau= \frac{4mz_T}{S_0N^2\hslash\,(1+p)^2\left(1-\displaystyle\frac{1}{2k_0z_t}\right)}\cdot\exp\!{\left(\frac{4k_0z_T}{3}\right)}
\label{eq:lifetime}
\end{equation}
where $z_T\!=\!-\varepsilon_0/eE$ is the outer classical turning radius, $\varepsilon_0$ is the binding energy, $p$ is a polarization factor of the ionic wave function ($p\!=\!0.0126$), $k_0^2\!=\!2m(-\varepsilon_0)/\hslash^2$, $S_{\!0}$ is a spectroscopy coefficient ($S_{\!0}\!=\!0.783\,(5)$ \cite{Keating}), and the normalization constant $N$ is given by $N=[2k_0(k_0+\alpha)(2k_0+\alpha)]^{1/2}\!/\alpha$ where $\alpha$ is a parameter for the ionic potential function \cite{Tietz} ($\alpha=3.806\!\cdot\!10^{10}\;\text{m}^{-1}$). However, it is more common to parameterize the decay time taking into account some approximations according to experimental results \cite{Stinson}. Thus, lifetime is reduce from \qqref{eq:lifetime} to:
\begin{equation}
\tau=\frac{A_1}{E}\cdot\exp{\left(\frac{A_2}{E}\right)}
\label{eq:lifetime_P3}
\end{equation}
where $A_1$ and $A_2$ are functions of binding energy experimentally determined \cite{Jason,Keating} ( \tabref{table:stripping_lifetime}).
\begin{table}[h] \scriptsize
	\centering
	\caption{Lifetime parameters for $H^-$}
	\begin{tabular}{|c|c|c|}\hline
		Parameter				& Jason \textit{et al.} \cite{Jason} &  Keating \textit{et al.} \cite{Keating} \\ \hline
		$A_1\;(10^{-6}$ s V/m)	& $2.47\,(10)$ 	& $3.073\,(10)$ \\ \hline
		$A_2\;(10^{9}$ V/m)		& $4.49\,(10)$	& $4.414\,(10)$	\\ \hline
	\end{tabular}
	\label{table:stripping_lifetime}
\end{table}

\subsection{1D synchrotron radiation model}  \label{ssec:csr}
Coherent synchrotron radiation (CSR) can greatly affect the quality of an electron beam traveling through a bending magnet. It is an important effect for FEL drivers due to the use of chicane bunch compressors. CSR in such a system can lead to increased energy spread, emittance dilution, and micro-bunching instability \cite{saldin2002}, \cite{heifets2002}. \opalt includes two 1D models to simulate CSR, \cite{Saldin1997} and \cite{mitchell2013}. In both, the transient wakefields which occur at the entrance and exit of a bend are included and in principle the two methods will give the same results. However, the second, more recent method \cite{mitchell2013} is less numerically noisy.

\subsection{Multi-Objective Optimization} \label{ssec:moo}
Optimization methods deal with finding a feasible set of solutions
corresponding to extreme values of some specific criteria. Problems
consisting of more than one criterion are called {\it multi-objective
optimization problems} (MOOP).

As with single-objective optimization problems,
MOOPs consist of a solution vector and optionally a
number of equality and inequality constraints. Formally, a general
MOOP has the form

\begin{align}
  \mathrm{min}          \quad & \quad f_m(\mathbf{x}),                                & m &= \{1, \ldots, M\} \\
  \mathrm{subject\, to} \quad & \quad g_j(\mathbf{x}) \geq 0,                         & j &= \{1, \ldots, J\} \\
  \quad                       & \quad  x_i^L \leq \mathbf{x}=x_i \leq x_i^U,& i &= \{0, \ldots, n \}.
\end{align}

The $M$ objectives $f_m(\mathbf{x})$ are minimized, subject to $J$
inequality constraints $g_j(\mathbf{x})$.
An $n$-vector $\mathbf{x}$ contains all the
design variables with appropriate lower $x_i^L$ and upper bounds $x_i^U$, constraining
the design space.

In most MOOPs conflicting objectives are encountered,
which complicates the concept of optimality.
One popular approach to tackle MOOPs are evolutionary algorithms~\cite{deb:09}.


In \opal one of the standard evolutionary algorithm,
the Non-dominated Sorting Genetic Algorithm-II~\cite{Deb00afast} has been implemented~\cite{Neveu:2013ues, ineichen2013}
with the PISA library~\cite{PISA}.

The implementation is based on a master/slave paradigm, employing several masters and groups of workers to prevent communication hotspots at master processes.
In addition, we exploit information about the underlying network topology when placing master processes and assigning roles.
This allows computationally efficient optimization runs with a large number of cores.

The following is an example of a MOOP definition and \texttt{OPTIMIZE} command in \opal:
\begin{verbatim}
phi0:  DVAR, VARIABLE   = "PHI0",
             LOWERBOUND = -180,
             UPPERBOUND =  180;
theta: DVAR, VARIABLE   = "THETA",
             LOWERBOUND = 0,
             UPPERBOUND = 45;
             
energy: OBJECTIVE,
   EXPR = "-statVariableAt('energy',600.0)";
        
opt: OPTIMIZE, INPUT   = input.tmpl,
     OBJECTIVES        = {energy},
     DVARS             = {phi0, theta},
     INITIALPOPULATION = 10,
     MAXGENERATIONS    = 20,
     NUM_IND_GEN       = 10,
     NUM_MASTERS       = 1,
     NUM_COWORKERS     = 1;
\end{verbatim}
Here the \texttt{OPTIMIZE} command specifies an optimization for two design variables (DVARs) and one objective, the maximum energy at a certain location.
The DVARs need to be specified in the templated input file.
The size of the initial population, number of individuals per generation and the number of generations is specified, as well as the number of masters and slaves.

\subsection{Sampler}
The sampler is a new feature in \opal that is based
on the implementation of the multi-objective optimization.\
Therefore, the syntax of design variables (DVARs) in the
input command follows the very same notation.\

However, instead of running an optimization, each individual
defining a single simulation is created by sampling the
DVARs according to some probability distribution.\
Currently, the sampler supports following methods:
\begin{itemize}
    \item \texttt{FROMFILE},
    \item \texttt{UNIFORM},
    \item \texttt{UNIFORM\_INT},
    \item \texttt{GAUSSIAN},
    \item \texttt{LATIN\_HYPERCUBE}.
\end{itemize}
An up-to-date version is also found in the manual.\
Although only specialized implementations of the uniform and
normal distribution are provided, a user is able to sample any
kind of distribution thanks to the attribute \texttt{FROMFILE}.\
Such a file may contain all or only a subset of DVARs
in a column-wise fashion where the header specifies the
DVAR names.\ These have to agree with the string
used in \texttt{VARIABLE} of the \texttt{DVAR} command.\ In this
case the \texttt{LOWERBOUND} and \texttt{UPPERBOUND} of
the corresponding variable(s) are ignored.\ A distribution may therefore
be generated using a third-party library and provided to the sampler.

Depending on the selected distribution type the \texttt{SAMPLING}
command accepts different attributes.\ When reading the samples
of a DVAR from a file it requires the flag \texttt{FNAME}.\ All
others take a \texttt{SEED}, the number of
samples \texttt{N} and a boolean \texttt{RANDOM}.\ In non-random mode a sequence of
DVAR inputs is generated following the underlying distribution.\  Examples of two
sampling types are given below:

\begin{verbatim}
SM1: SAMPLING, VARIABLE="dvar",
               TYPE="FROMFILE",
               FNAME="/path/to/file/fname.dat";
\end{verbatim}

\begin{verbatim}
SM2: SAMPLING, VARIABLE="dvar",
               TYPE="GAUSSIAN",
               SEED=122,
               N=1000;
\end{verbatim}

The last ingredient to run the sampler is the \texttt{SAMPLE} command.\
It is similar to the optimizer.\ Here we just highlight a few special
features.\ Each \opal{} simulation spawned by the sampler generates output files that
may not be requested for post-processing.\ Furthermore, symbolic links that point to
input files like field maps etc.\ can be deleted without hesitation.\ For this purpose
the \texttt{SAMPLER} command provides the attribute \texttt{KEEP} that takes a list of
file extensions.\ Another important boolean attribute \texttt{RASTER} specifies the way to
combine the univariate sample distributions.\ Assuming $d$ design variables of $N_1, ..., N_d$ samples.\
In raster mode all combinations are made and, thus, the final number of \opal{} simulations is $N_1\times ... \times N_d$.\ In regular mode, i.e. \texttt{RASTER=FALSE}, the $i$-th simulation is created by the $i$-th sample
point of every DVAR, ending up with $\min_{i=1,...,d}{N_i}$ \opal{} runs.\


%
\section{Benchmarks}  \label{sec:benchm}

\subsection{FFA single particle tracking}  \label{ssec:benchffag}

The beam orbit in an FFA accelerator moves radially with momentum, as in a cyclotron. Simulation codes which assume a central orbit independent of momentum are unsuitable for studying FFAs as they do not reproduce the correct dynamics. \opal is one of only a few simulation codes which remove the constraint of the existence of this central orbit. 

A simulation code benchmarking was undertaken as part of an experimental collaboration using the 150 MeV FFA accelerator at Kyoto University Research Reactor Institute (KURRI). The general aim of the collaboration is to progress toward high intensity operation of this machine, the relevant parameters of which can be found in Ref.~\cite{ref_sheehy14}. The simulation campaign has been established alongside the experiments to provide reliable FFA modelling tools and to complement the experiments. Other codes used by collaboration members for the low intensity benchmarking include Zgoubi~\cite{ref_ZGOUBI}, SCODE~\cite{ref_SCODE}, MAUS~\cite{ref_MAUS} and EARLIETIMES~\cite{ref_EARLIETIMES}.

In order to provide a realistic benchmark lattice, we use the 3D magnetic field map of one DFD triplet computed with TOSCA. This is required for the 150 MeV KURRI-FFA as the magnets have a relatively large gap height and therefore a large fringe field extent. The transverse tunes and revolution frequencies are not accurately reproduced using a hard edge model.

The TOSCA field map is computed with grid points in a cylindrical coordinate system typically every 1\,cm. In the vertical direction, one grid layer is provided above and below the mid-plane, which gives the field gradient in that direction. The 3D field components at an arbitrary space coordinate are interpolated with the neighbouring grid points linearly (SCODE and EARLIETIMES) or with higher order interpolation (\opal, MAUS and Zgoubi).

\begin{figure}[htbp]
\begin{center}
\includegraphics[width=0.8\linewidth]{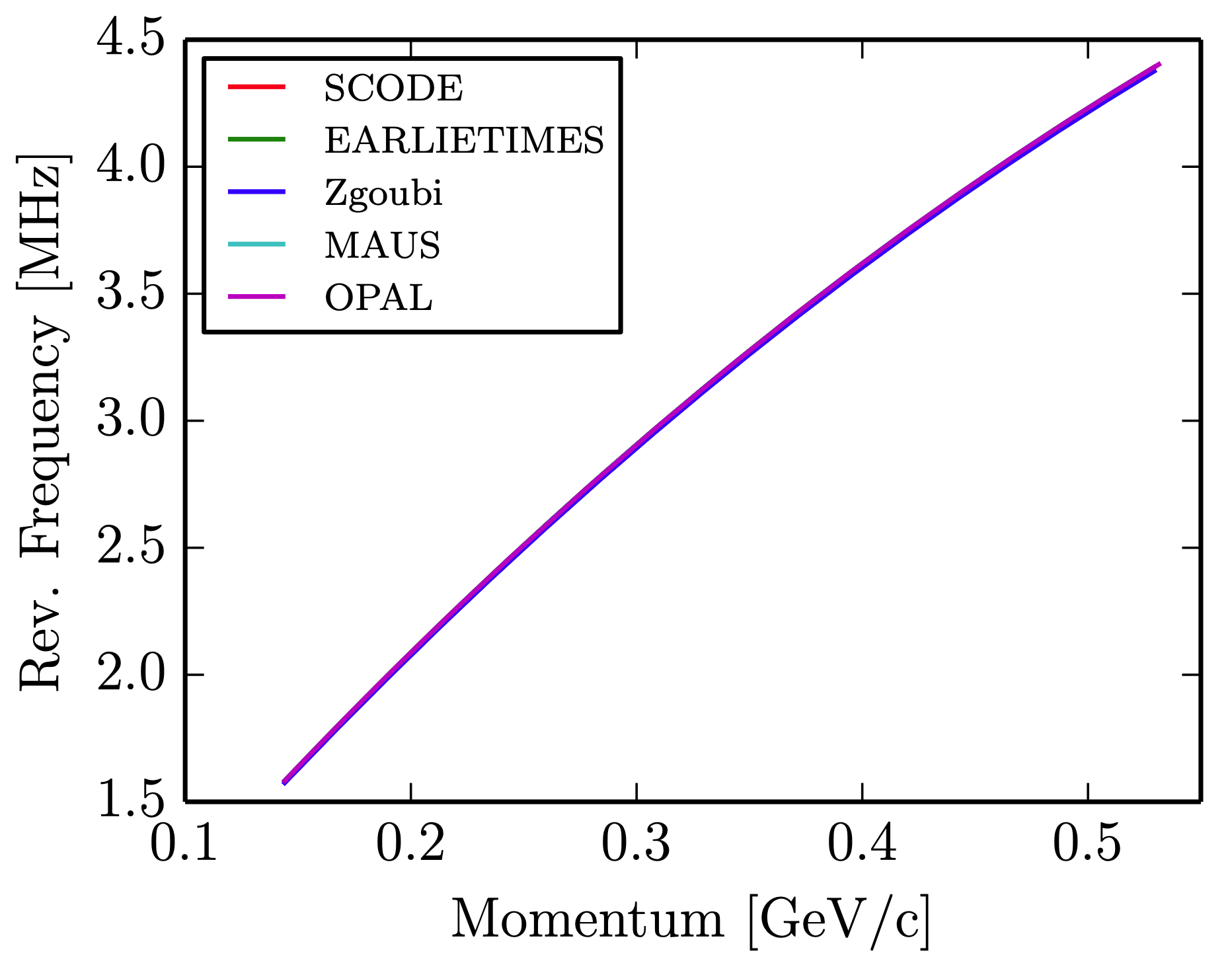}
\caption{Revolution frequency vs momentum for kinetic energies in the range from 11 to 140\,MeV \cite{bencheff}.}
\label{FFAbench_revf}
\end{center}
\end{figure}

The first step in simulation of the FFA is to establish the location of the closed orbits at each momentum throughout the 11 to 150\,MeV energy range. Combining the 12 DFD cells of the ring, the closed orbits are established with single particle tracking and this provides the revolution frequency with momentum, shown in \figref{FFAbench_revf}. This then allows us to calculated the optics properties including the betatron tune.

For the betatron tune comparison, the TOSCA field map was calculated with realistic excitation currents that can later be benchmarked with experiment. That is, 810~A for the F magnet at 1020~A for the D magnets. The ideal lattice case was assumed here, neglecting at this stage the knowledge that there is a large distortion of the closed orbit in the experiment. The integration step in each code was optimised until the betatron tune with momentum results became independent of step size. 

\begin{figure}[htbp]
\begin{center}
\includegraphics[width=0.8\linewidth]{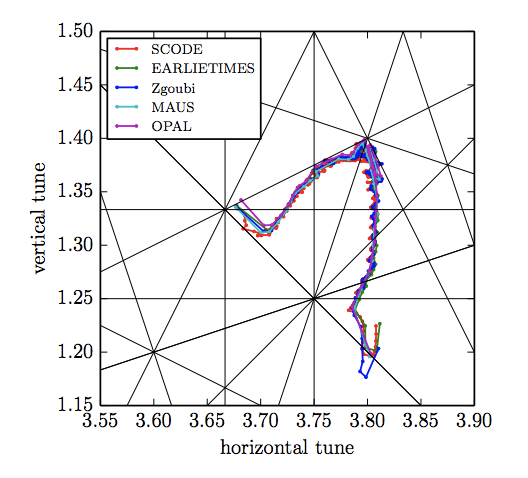}
\caption{Betatron tune from 11 to 139 MeV of the 150 MeV FFA at KURRI,
\cite{bencheff} calculated with SCODE, EARLIETIMES, Zgoubi, MAUS and \opal.}
\label{FFAbench_tunes}
\end{center}
\end{figure}

The results are shown in Fig.~\ref{FFAbench_tunes}, where the general agreement between the different codes is excellent. We note there are slight deviations between codes at the start and end of the momentum range, due to poor interpolation at the first and last grid point of the 3D map.

Having established these basic parameters, the simulation benchmarking will next move toward modelling realistic acceleration, tracking and effects with space charge.

\subsection{Space Charge Models}  \label{ssec:benchsc}
In a recent publication \cite{winklehner:spiral}, the 
smooth aggregation AMG solver (SAAMG) was benchmarked
against the analytical solution of a quasi-infinite beam in a conducting beam
pipe, both centered and off-centered. Furthermore, both SAAMG and FFT solver were
compared with the experimental results of injecting a DC ion beam into a cyclotron using
a so-called spiral inflector. All benchmarks yielded good agreement.
The following two subsections summarize these results.
\subsubsection{Analytical model}
Solving the Poisson equation for an infinitely long beam (no longitudinal space-charge forces) with
uniform density and hard cut-off at the beam radius $r_b$ we find the potential inside (superscript ``in'') and outside (superscript ``out'') of $r_b$ to be
\begin{equation}
\label{eqn:epot}
  \begin{aligned}
    \phi^{\mathrm{in}} &=
    \tilde{\phi}\left[1 + \ln\left(\frac{\chi}{r_p^2 r_b^2}\right)
    - \frac{(x-\xi)^2 + y^2}{r_b^2}\right] , \\ 
    \phi^{\mathrm{out}} &= 
    \tilde{\phi}\cdot\ln \left[ \frac{\chi}{r_p^2(x-\xi)^2 + r_p^2 y^2}\right].
   \end{aligned}
\end{equation}   
Here
\begin{equation*}
\label{eqn:deltaphi_noneut}
\tilde{\phi} = \frac{I}{4 \pi \epsilon_0 v_b}
\end{equation*}
(where $I$ is the beam current and $v_b$ the beam velocity), and
\begin{equation*}
\chi = \xi^2y^2+(\xi x - r_p^2)^2.
\end{equation*}
with $r_p$ the beam pipe radius and $\epsilon$ the distance from beam center to pipe center.
The electric field is calculated from the potential using
\begin{equation}
  \label{eqn:efield}
	\mathbf{E}^{\mathrm{in, out}} = -\nabla \phi^{\mathrm{in, out}}.
\end{equation}	

\begin{figure}[t!]
\centering
	\includegraphics[width=0.45\textwidth]
	                {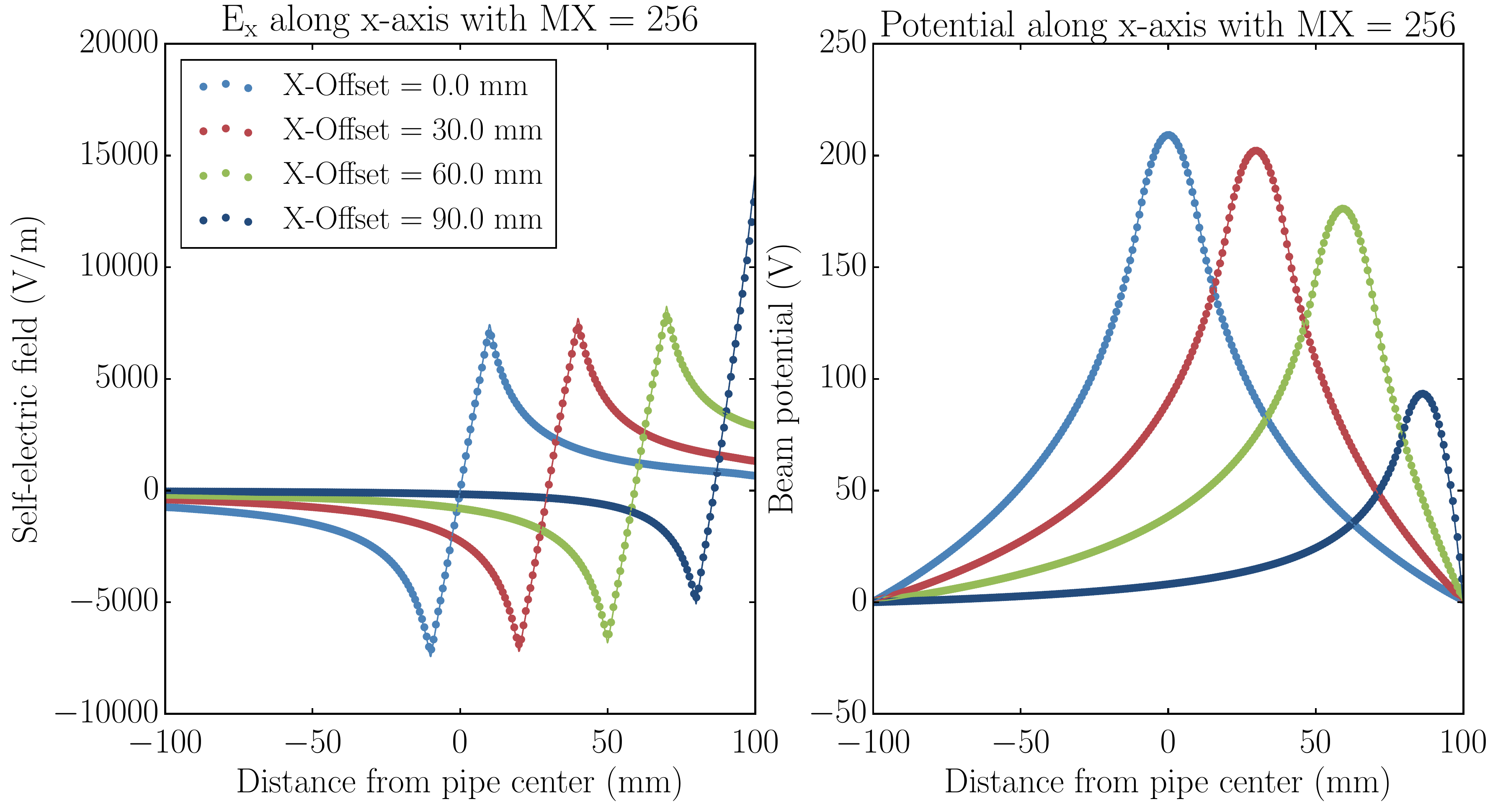}
	\caption{$\phi$ and $E_x$ along the x-axis for different offsets $\xi$. 
	         Dots are values calculated by \opal, while solid lines are the 
	         analytical solution. Excellent agreement can be seen with
	         $0.01 < \chi^2_{\mathrm{red}} < 0.02$ for the potential and 
	         $0.03 < \chi^2_{\mathrm{red}} < 1.3$ for the field. From \cite{winklehner:spiral}}
	\label{fig:offset_variation}
\end{figure}

The results of \opal calculation using the SAAMG field solver are plotted together with the 
analytical prediction from \eqref{eqn:epot} and \eqref{eqn:efield} in \figref{fig:offset_variation}.
Here the mesh size was 256 x 128 x 512 cells in x, y, and z, respectively and the offset $\epsilon$
was varied. Excellent agreement can be seen.

\subsubsection{Benchmark against experiment}
\begin{figure}[!b]
\centering
	\includegraphics[scale=0.3]  
	                {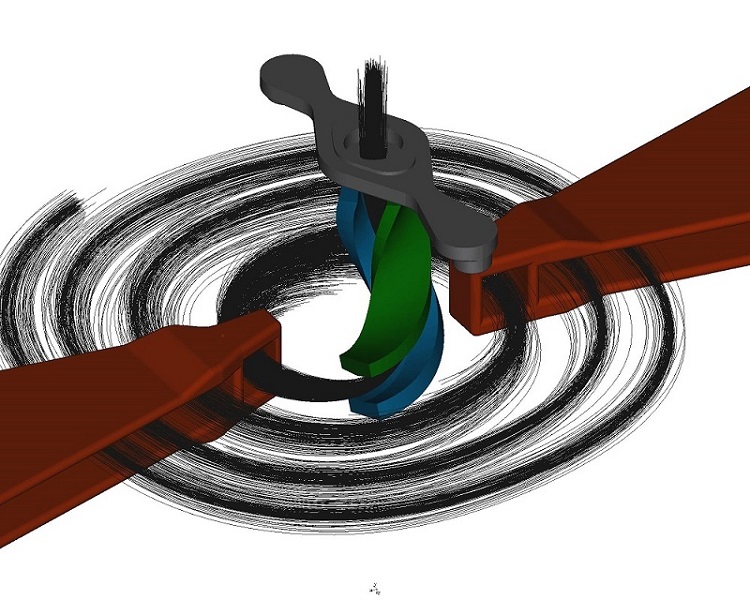}
	\caption{A spiral inflector with particle trajectories from
			 an \opal simulation. From \cite{winklehner:spiral}}
	\label{fig:spiral_schematic}
\end{figure}

A spiral inflector is an electrostatic device typically used to bring an ion beam from the axial
direction into the cyclotron mid-plane to be subsequently accelerated. The inflector
is shaped in such a way that the electric field complements the cyclotron's main magnetic field, 
and the combination of both fields forces the particles on the desired path. 
This process is shown in \figref{fig:spiral_schematic}. 

As reported in \cite{winklehner:spiral}, \opal was updated with the capability of accurately
describing such a spiral inflector within the \opalcycl flavor. This update included the
geometry update described in \secref{ssec:geom}, modifications to \opalcycl, and modifications 
to the SAAMG solver. In 2013/14, measurements of proton and \htp beams moving through a spiral
inflector like the one depicted in \figref{fig:spiral_schematic} were performed at Best Cyclotron
Systems, Inc. in Vancouver \cite{winklehner:bcs_tests, winklehner:spiral} and compared to \opal
simulations. Much care was taken to obtain realistic initial conditions for the inflector 
simulations that matched the 
experimentally observed beam parameters before the injection well.
Generally, good agreement was found between FFT solver, SAAMG solver and \opal, when comparing
the results of radial probe measurements after the spiral inflector.
Noticeable differences between FFT and SAAMG were only seen for beam currents significantly
higher than the experimental ones.

\subsubsection{Photoinjector}
\begin{figure}[ht]
    \centering
    \includegraphics[width=\linewidth]{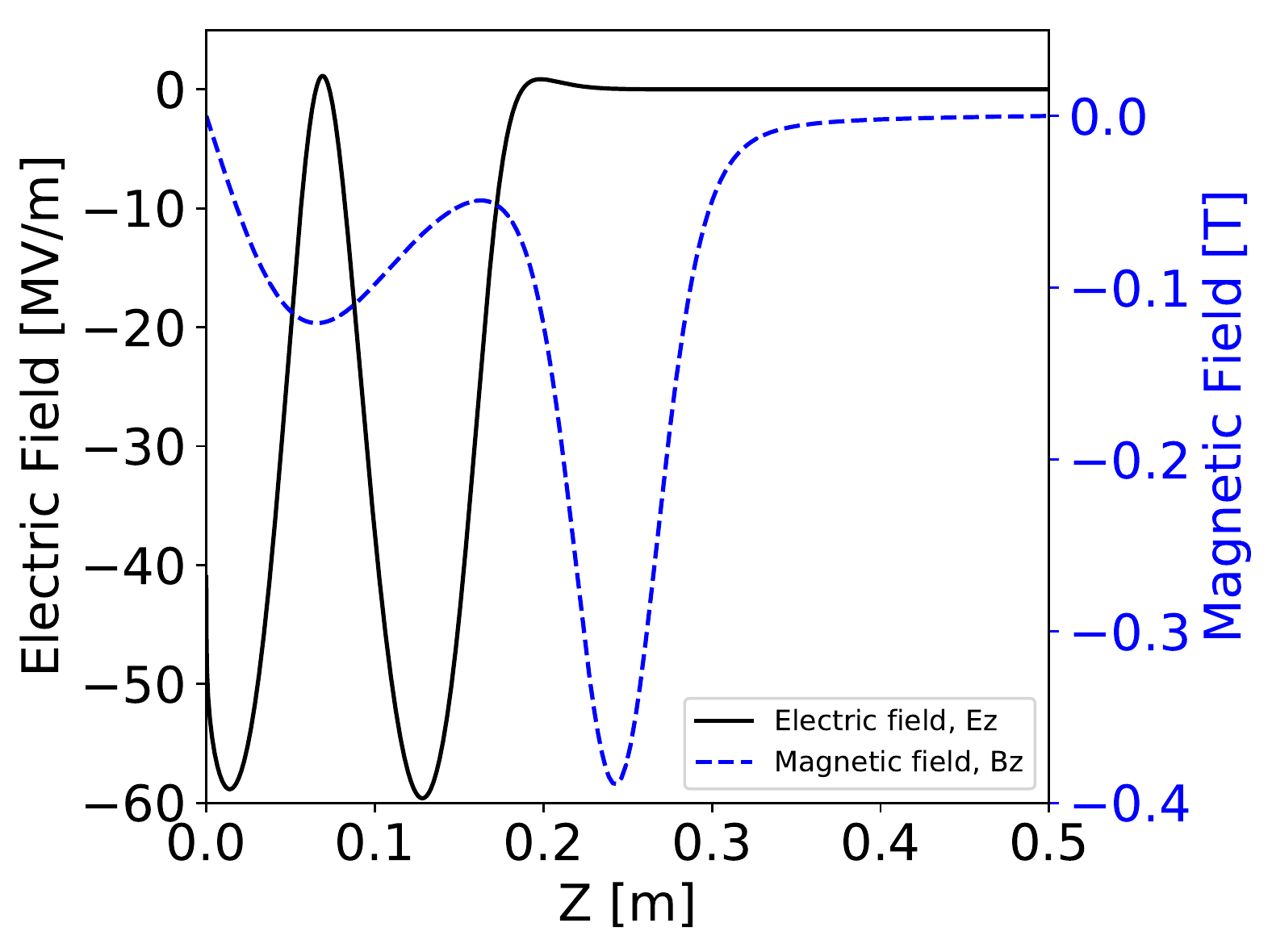}
    \caption{Magnetic and electric fields in the gun.}
    \label{fig:EMawagun}
\end{figure}

The space charge algorithms in ASTRA, GPT, and \opalt, were compared using a photoinjector (gun) at the Argonne Wakefield Accelerator Facility (AWA). 
This work was originally presented here~\cite{napac16}.
The simulation model included a 1.5 cell copper standing-wave cavity at \SI{1.3}{GHz}, with bucking, focusing, and matching solenoids. 
The rf gun and solenoid fields seen by the beam are shown in ~\figref{fig:EMawagun}. 
The simulation parameters were chosen to approximately generate \SI{1}{um} at \SI{1}{nC} case. 
The initial beam parameters were based on gun operations at PITZ~\cite{pitz}, 
due to the similarities between the PITZ and AWA rf guns. 
The PITZ parameters came close to achieving the \SI{1}{um} target without any optimization. 
A coarse 1D minimization of the emittance was done to adjust the laser radius. 
The resulting minimum emittance was \SI{1.16}{um}.

The initial bunch distribution parameters as well as the on-axis gun gradient 
($E_z$) and magnetic field ($B_z$) are listed in Table~\ref{tab:awa_gunparams}. 
The rf gun and solenoid field maps were generated with the SUPERFISH/POISSON codes~\cite{superfish,poisson}. 
The gradient was chosen to match typical operations at PITZ~\cite{pitz} and the AWA. 
Note that the codes use various methods to model the rf and magnetic fields, SC, and image charge.
\begin{table}
    \centering
        \caption{Input Parameters for AWA gun}
    \begin{tabular}{ll}
    \toprule
        \textbf{Parameter} &  \textbf{Value} \\
        \midrule
         Charge & \SI{1}{nC} \\
         Laser Radius & \SI{0.75}{mm} \\
         Rise and Fall Time & \SI{6}{ps} \\
         Full Width Half Maximum & \SI{20}{ps} \\
         Phase & On Crest (Max Energy) \\
         kE at Cathode & \SI{0.55}{eV} \\
         Gradient on Cathode & \SI{60}{MV/m} \\
         Buck and Focusing Solenoid & \SI{-0.12}{T} \\
         Matching Solenoid & \SI{-0.389}{T}\\
         \bottomrule
    \end{tabular}
    \label{tab:awa_gunparams}
\end{table}

In ASTRA, the radial and longitudinal number of cells composing the mesh were 
set to $N_r=32$ and $N_z=64$ respectively with 100,000 particles. 
The image charge close to the cathode was accounted for until the bunch reached \SI{9.7}{cm} from the cathode surface.
GPT read in the 2D electric and magnetic field files, and used a square 3D adaptive SC mesh of $N_x=N_y=N_z=46$ with 100,000 particles. 
The image charge in GPT is calculated until the distance between the beam and cathode is longer than the mesh box.

\begin{figure}
    \centering
    \includegraphics[width=\linewidth]{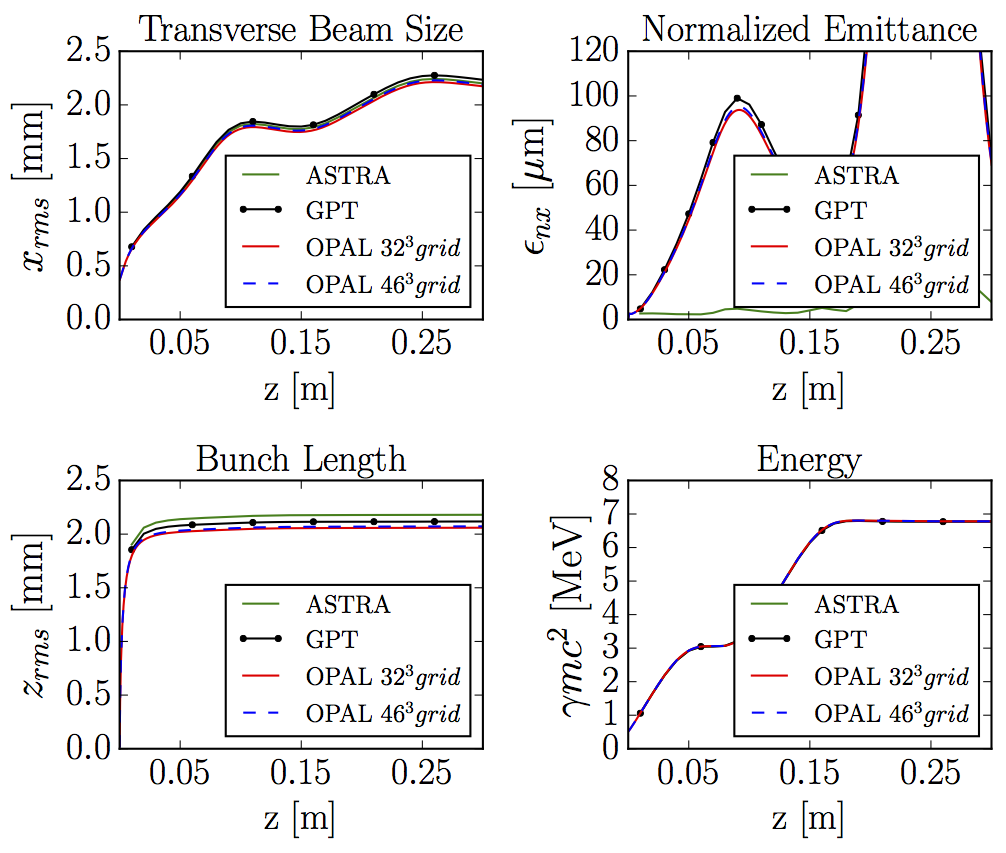}
    \vspace{-2em}
    \caption{Beam envelopes in the gun.}
    \label{fig:awa_ingun}
\end{figure}

\begin{figure}
    \centering
    \includegraphics[width=\linewidth]{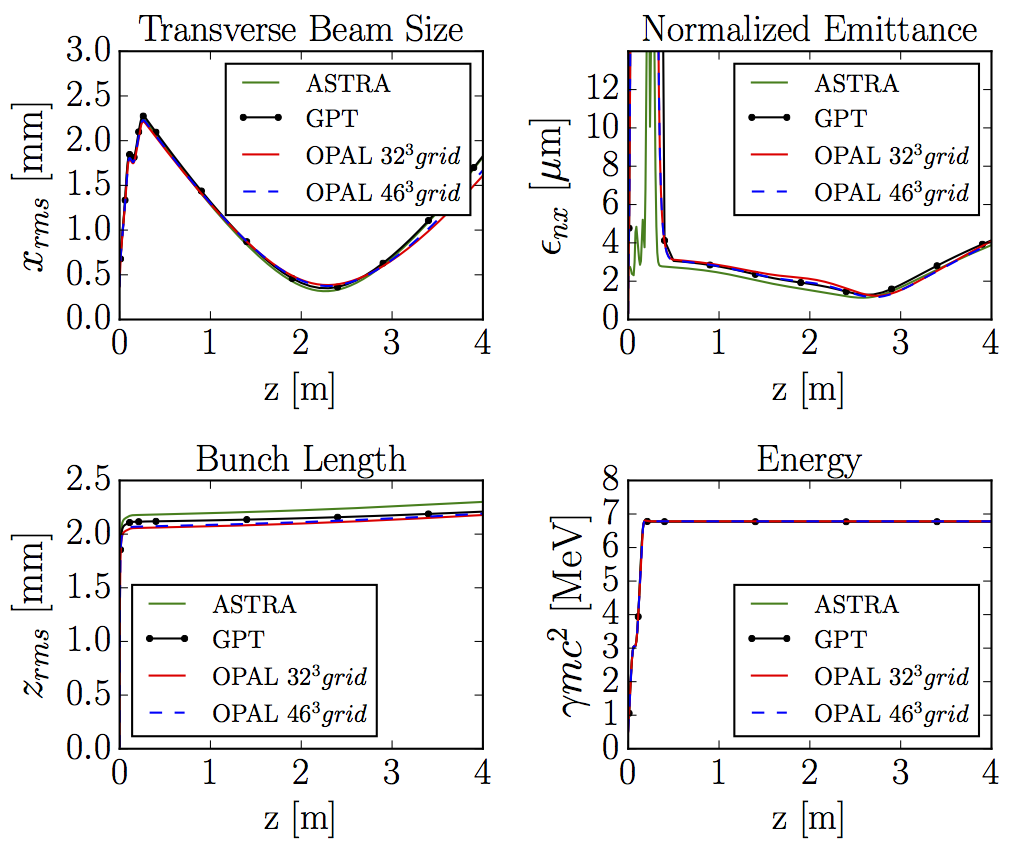}
    \vspace{-2em}
    \caption{Beam envelopes in gun and drift.}
    \label{fig:awa_indrift}
\end{figure}

\opalt also read in the field maps, and used a block structured equidistant SC mesh. 
Two square mesh sizes were run, $N_x=N_y=N_z=32$ and $N_x=N_y=N_z=46$, both with 1 million particles. 
The later grid size was chosen to match GPT, and the number of particles was chosen to 
ensure at least 10 particles per grid cell.
The user can define how many time steps the image charge calculation should be carried out. 
The default value in \opalt is 200 time steps.
In general, the simulation results are in reasonable agreement and within expectations based on previous benchmarks~\cite{Limborg}. 
See Fig.~\ref{fig:awa_ingun}-\ref{fig:awa_indrift} for beam envelopes in the gun and drift. 
The apparent disagreement of emittance between ASTRA and the other two codes 
in the gun is because the former removes the angular momentum induced by the solenoid, 
while the later two codes do not. 
After the beam exits the solenoid, the emittance results are in good agreement, as shown in Fig.~\ref{fig:awa_indrift}.

\subsection{Particle Matter Interaction Model} \label{ssec:benchpm}


\label{sec:Deg}

For this benchmark, in the realm of proton therapy beam simulations, we consider as reference
FLUKA, a general Monte Carlo simulation package for calculations of particle transport and interactions with matter \cite{Fluka1}.

\begin{figure}[ht]
\centering
\includegraphics[scale=0.26, keepaspectratio=true]{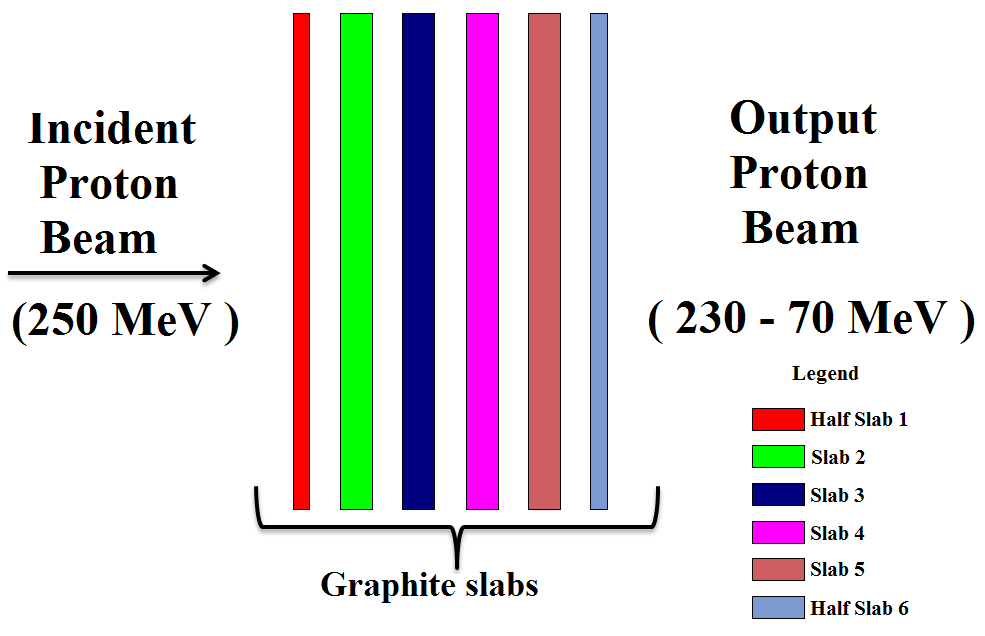}
    \caption{Schematic representation of the multi-slabs geometry implemented in FLUKA}
    \label{fig:Deg_slab}
\end{figure} 

We will consider two degrader geometries, a realistic multi-slabs geometry (see \figref{fig:Deg_slab}) and an equivalent single slab geometry. 
Five slab thicknesses are considered that roughly correspond to five degrader energy settings: 230, 200, 150, 100 and 70~MeV.
We also use the initial FLUKA input beam in \opal in order to recreate the same initial condition. In case of the multi-slabs geometry, after each slab, a monitor is placed for recording the beam phase space.
The relevant \opal\ elements are sketched below.
\begin{verbatim}
DEGPHYS_Slab1 : SURFACEPHYSICS,
                TYPE       = "DEGRADER", 
                MATERIAL   = "GraphiteR6710";

Wedge1: DEGRADER, L        = 0.00255124,
                  ZSIZE    = 0.00255124,
                  OUTFN    = "Wedge1.h5", 
            SURFACEPHYSICS = DEGPHYS_Slab1, 
                  ELEMEDGE = 0.02;
        
Deg_D1: DRIFT, L           = 0.0360075,
               ELEMEDGE    = 0.02255;

MDegD1: MONITOR, OUTFN     = "MDegD1.h5", 
                 ELEMEDGE  = 0.04255;
\end{verbatim}

\subsubsection{Multi-slabs}

Different from FLUKA, the time step is an additional variable in the \opal simulation. 
Keeping the degrader setting for 230 MeV fixed, the time step influence on the final results has been studied and the results are 
shown in \tabref{table:MS_230_TS}.

\begin{table}[ht]\scriptsize
\centering
 	\begin{tabular}{|c|c|c|c|c|}
		\hline
		\textbf{Position} & \textbf{1 ps} 	& \textbf{3 ps} & \textbf{5 ps}  & \textbf{7 ps} \\
		\hline
		HSlab1 		  &	247.88	(-0.04) &	247.67	(0.04) &	248.19	(-0.17) &	248.22	(-0.18) \\
		\hline
		Slab2                 &  244.48	(-0.04) &	244.03	(0.14) &	245.13	(-0.31) & 245.26	(-0.36) \\
		\hline
		Slab3 		  & 241.06	(0.001) &	240.38	(0.28) &	242.08	(-0.42) &	242.49	(-0.59) \\
		\hline
		Slab4 		& 237.62		(-0.02) &	237.07	(0.21) &	238.55	(-0.41) &	238.72	(-0.48) \\
		\hline
		Slab5 		& 234.14		(-0.01) &	233.75	(0.16) &	234.97	(-0.36) &	235.24	(-0.47) \\
		\hline
		HSlab6		& 232.39		(-0.01) &	231.92	(0.20) &	233.12	(-0.32) &	233.49	(-0.48) \\
	  	\hline
		\end{tabular}
\caption{Degrader setting for 230 MeV: mean energy values in MeV after each slab of graphite for various time steps.
In parenthesis, the percentage discrepancy with FLUKA results.}
\label{table:MS_230_TS}
\end{table}
The discrepancy between \opal and FLUKA is monotonic decaying as expected and below 0.5\% for all time steps considered.


\begin{table}[ht]\scriptsize
\centering
 	\begin{tabular}{|c|c|c|c|}
		\hline
		\textbf{Position} & \textbf{230 MeV} & \textbf{200 MeV} & \textbf{150 MeV} \\
		\hline
	  	HSlab1	& 247.88	(0.00)	 	& 245.11	(-0.03)  & 241.09	(-0.01) \\
		\hline
		Slab2	& 244.48	(0.00)	 	& 236.18	(-0.02)  & 223.55	(0.01)  \\
		\hline
		Slab3	& 241.06	(0.00)   	& 227.01	(-0.03)  & 205.14	(0.04) \\
		\hline
		Slab4	& 237.62	(0.00) 	& 217.58	(-0.02)  & 185.61	(0.07) \\
		\hline
		Slab5	& 234.14	(0.00) 	& 207.88	(-0.01)  & 164.62	(0.10) \\
		\hline
		HSlab6	& 232.39	(0.00) 	& 202.92	(-0.01)  & 153.41	(0.13) \\
	  	\hline
		\end{tabular}
\caption{Mean energy values in MeV after each slab of graphite in \opal using a 1~ps time step. In parenthesis, the percentage discrepancy with FLUKA results.}
\label{table:OPAL_FLUKA_1ps}
\end{table}

Three out of five energy settings can be analysed with the multi-slabs geometry where between each slab a monitor is placed, see \tabref{table:OPAL_FLUKA_1ps}.
With the increase in the thickness of the graphite slabs, the drift space in between becomes too short for placing a monitor. In \opal, the monitors 
do not have thickness but they need a region of $\pm$5~mm around for being active, otherwise they are ignored and data not stored. 
This is the case for the 100 and 70~MeV setting. 


\subsubsection{Single slab}

To avoid the problem with placing the monitor when the slabs are close, another geometry has been implemented in \opal and FLUKA. The six slabs are replaced with a single slab of equivalent thickness. A monitor (or the equivalent USRBNX card in FLUKA) is placed right after this slab for recording the energy reduction.

In FLUKA, simulations with and without inelastic scattering have been performed. The output data for the energy loss are 
used as reference for the comparison with \opal.

In \opal, the influence of the time step tested for each energy setting. The results and comparison with FLUKA is shown in \tabref{table:OPAL_FLUKA_Ene_SS}.

\begin{table}[ht]\tiny
\centering
 	\begin{tabular}{|c|c|c|c|c|c|}
	\hline
	\textbf{Time Step (ps)}  &	\textbf{230 MeV} & \textbf{200 MeV} & \textbf{150 MeV} & \textbf{100 MeV} & \textbf{70 MeV}\\
	\hline
				1	    &		 232.38	(0.00)  &	202.83	(0.04)   &	153.41	(0.13)  & 103.50	(0.33)	& 75.09	(0.61) \\
	\hline
				3	   &   	232.27	(0.05)  &  202.66	(0.12)   &	153.29	(0.21)   & 103.34 (0.48)	&  74.91	(0.85) \\
	\hline
				5	   &     	232.40	(-0.01) &	202.95	(-0.02)  & 153.64	(-0.02)   & 103.82 (0.02)	& 75.54	(0.01) \\
	\hline
				7	  &     	232.30	(0.03)   &	203.35	(-0.22)  &	153.88	(-0.18)   & 104.12 (-0.27) & 75.93	(-0.50) \\
	\hline
\end{tabular}
\caption{Mean energy values (MeV) after the single graphite slab in \opal for different time steps. In parenthesis, the percentage discrepancy with FLUKA.}
\label{table:OPAL_FLUKA_Ene_SS}
\end{table}

Even with this geometry the discrepancy between \opal and FLUKA is less than 1\% for all the energies. Different from the multi-slabs analysis, 5~ps seems to be the proper time step to set for this Monte Carlo simulation. 



Further comparisons between different Monte-Carlo codes and \opal for the degrader in proton therapy facilities are discussed in~\cite{RIZZOGLIO20181}.

%
%
%
%

\subsection{1D Synchrotron Radiation Model} \label{ssec:benchcsr}
\begin{figure}[ht!]
  \begin{center}
  \includegraphics[width=.7\columnwidth,angle=-90]{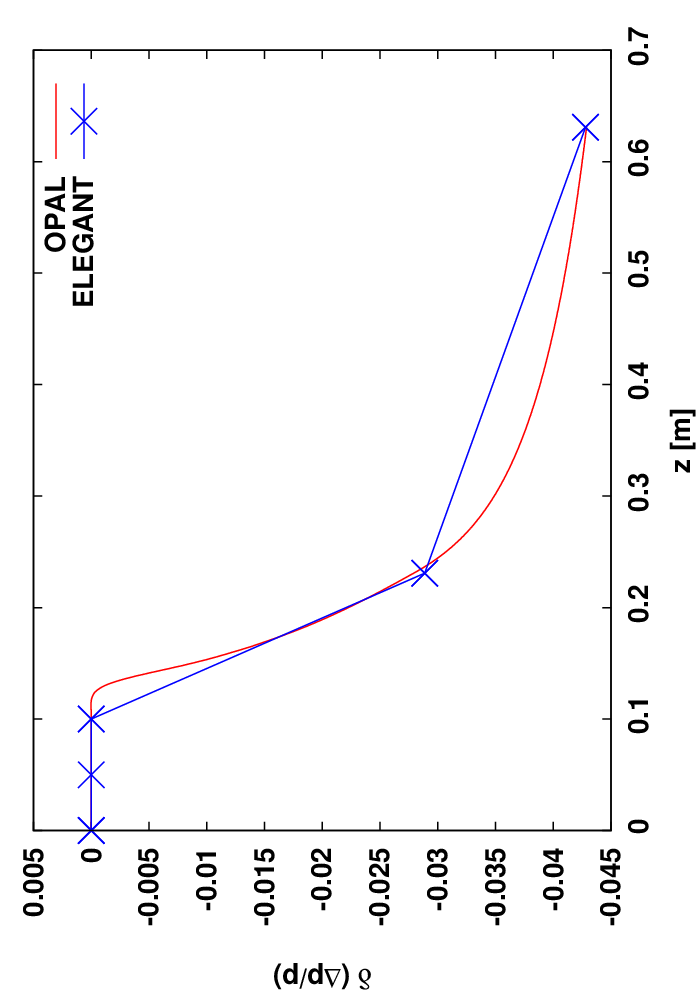}
    \end{center}
  \caption{Comparison of CSR routines in \opal and ELEGANT.}
  \label{fig:csr}
\end{figure}

We validated \opal CSR routine against the ELEGANT. In \figref{fig:csr}, we compare the $\delta = \Delta p/p$ obtained by \opal and ELEGANT by sending a 7 MeV electron bunch with 1 nC of charge through a beam line which is consisted of a 30 degree bend with 0.25 meter radius preceded by a 0.1 meter long drift and followed by a 0.4 meter long drift. Good agreement between \opal and ELEGANT can be achieved.

\begin{figure}
    \centering
    \includegraphics[width=0.7\columnwidth]{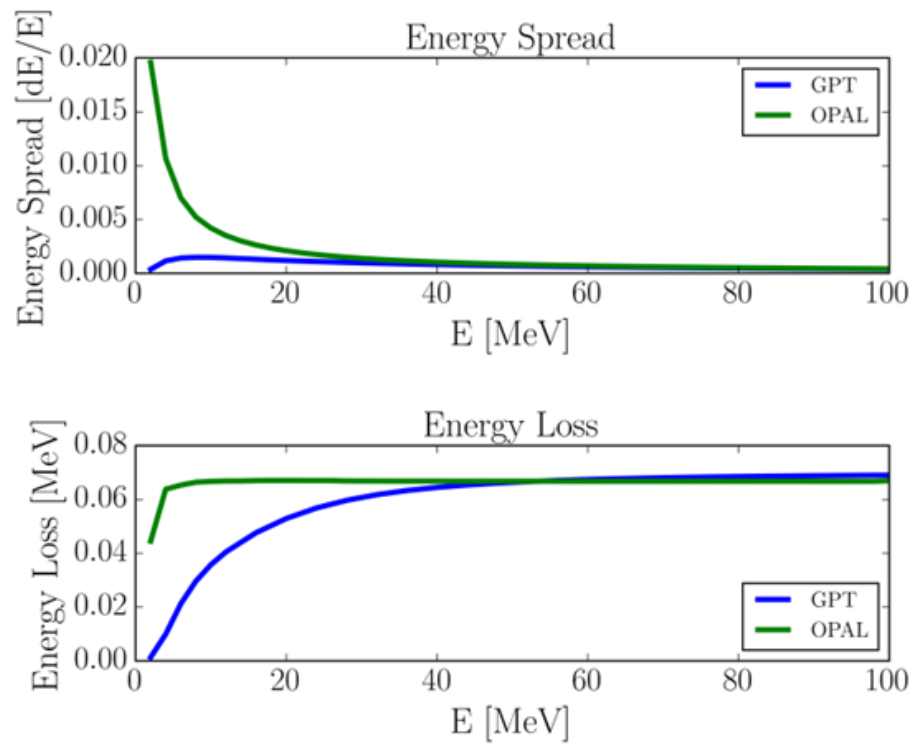}
    \caption{Comparison of CSR routines in \opalt and GPT.}
    \label{fig:gptcsr}
\end{figure}

A comparison with GPT and \opalt was also done.
Short monoenergetic Gaussian bunches with zero initial emittance were sent through a dipole.
The transverse beam size ($\sigma_x$, $\sigma_y$) was set to \SI{1}{mm}.
The bunch length was set to \SI{0.3}{mm}. The bending angle was fixed at $20^\circ$, 
and the beam energy was varied from $2-100$ MeV.
The CSR routine in GPT does not use the ultra-relativistic approximation ($\beta = 1$) and as a result, works at all energies.
Therefore, we expected the routines to match well at high energy and diverge at lower energy. Results of the CSR simulations are shown in Fig.~\ref{fig:gptcsr}.
As expected, the results between GPT and \opalt disagree at low energies.


%
\section{Applications} \label{sec:appl}

\subsection{The PSI high intensity cyclotrons} \label{ssec:benchhipa}


The High Intensity Proton Accelerator (HIPA) at PSI delivers a proton beam of 590~MeV energy at a current up to 2~mA (1.2~MW).
It contains two consecutive cyclotrons,
the 4-sector \textit{Injector-2} which increases the beam energy from 870~keV up to 72~MeV
and the 8-sector \textit{Ring} Cyclotron for the final acceleration to 590~MeV.

\subsubsection{Tune Calculation}

\begin{figure}[ht]
\centering
        \includegraphics[width=1.0\columnwidth]
                        {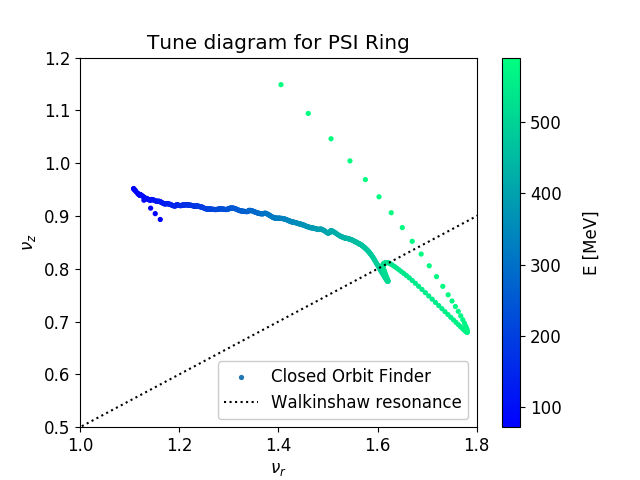}
        \caption{Tune diagram for the PSI Ring Cyclotron.}
        \label{fig:psiapp_tunes}
\end{figure}

\opalcycl has a closed orbit finder and tune calculation based on Gordon's algorithm \cite{Gordon:1984zz}.
An example for the Ring Cyclotron is given in \figref{fig:psiapp_tunes}.

\subsubsection{Trim Coils}

\begin{figure}[ht]
\centering
        \includegraphics[width=1.0\columnwidth]
                        {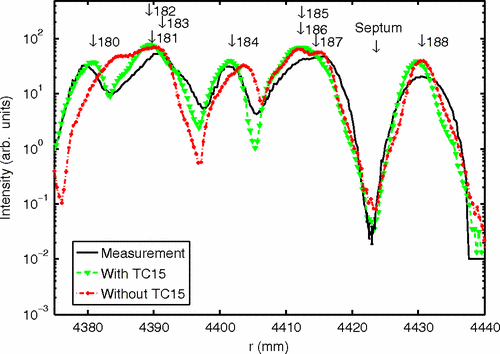}
        \caption{Radial beam profile with indicated turn numbers at extraction for a 2~mA beam. Taken from~\cite{PhysRevSTAB.14.054402}.}
        \label{fig:psiapp_extraction}
\end{figure}

The tune calculation was performed without trim coils and in this case the Wilkinshaw coupling resonance $\nu_r = 2 \nu_z$ is crossed four times (\figref{fig:psiapp_tunes}) resulting in large vertical beam losses.
In the Ring cyclotron the coupling resonance is avoided with a special trim coil, named TC15,
which was modeled in~\cite{PhysRevSTAB.14.054402}.
With this trim coil model the turn pattern at extraction could be matched very well with measurement, see \figref{fig:psiapp_extraction}.

The other 17~trim coils were modeled in~\cite{Frey:2019rmk} after a general trim coil model was added to \opalcycl.
With this model and the multi-objective optimisation (see Section \secref{ssec:moo}) all 182 turns in the Ring Cyclotron could be matched to measurements within 4.5~mm~\cite{Frey:2019rmk}.

\subsubsection{Injector-2 Collimators}

\begin{figure*}[t]
\centering
    \includegraphics[width=\textwidth]                {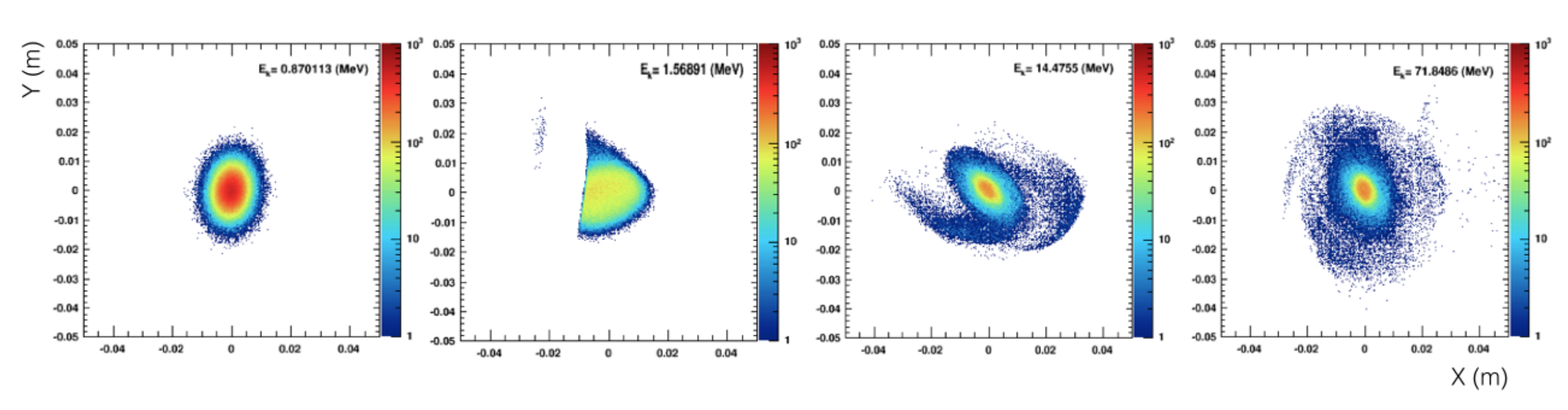}
    \caption{An example of progression of the distribution in the configuration space 
passed through collimators and accelerated to 72~MeV. Despite significant distribution
deformations due to large cuts, the tails ``wrap" around the spiralling centre of the
distribution leading to formation of a stable core with halo around. Taken from~\cite{kolano2017}.}
        \label{fig:psiapp_inj2beam}
\end{figure*}

To accurately describe the losses in a cyclotron,
the halo needs to be modelled very precisely.
In the Injector-2 cyclotron the incoming space charge dominated 11~mA beam with an energy of 970~keV is collimated after injection by 14~movable collimators in the central region.
This complicated geometry was successfully modeled in~\cite{kolano2017}.
An example of the beam distribution at various stages in the cyclotron is shown in \figref{fig:psiapp_inj2beam}.
The formation of a compact core with large halo around it can clearly be seen.
This follows predictions as described in~\cite{baartman-2013}, which states that {\it ``a non-matched non-circular bunch will match
itself after a number of turns (\dots) and the generated halo will depend upon the initial
mismatch"}.


\subsection{Argonne Wakefield Accelerator Facility}
The Argonne Wakefield Accelerator Facility (AWA), 
conducts accelerator R\&D on a variety of topics.
Recent experiments include emittance exchange~\cite{eex}, 
wakefield structure tests~\cite{xueying}, two beam acceleration (TBA)~\cite{tba}, 
and plasma wakefield acceleration~\cite{pwfa}.
As part of the the design efforts for future TBA experiments, 
\opalt was used to model and optimize several components of the high charge linac at the AWA.
A sample of this work is shown here.

\subsubsection{Solenoid Scans}
\begin{figure}
    \centering
    \includegraphics[width=0.45\linewidth]{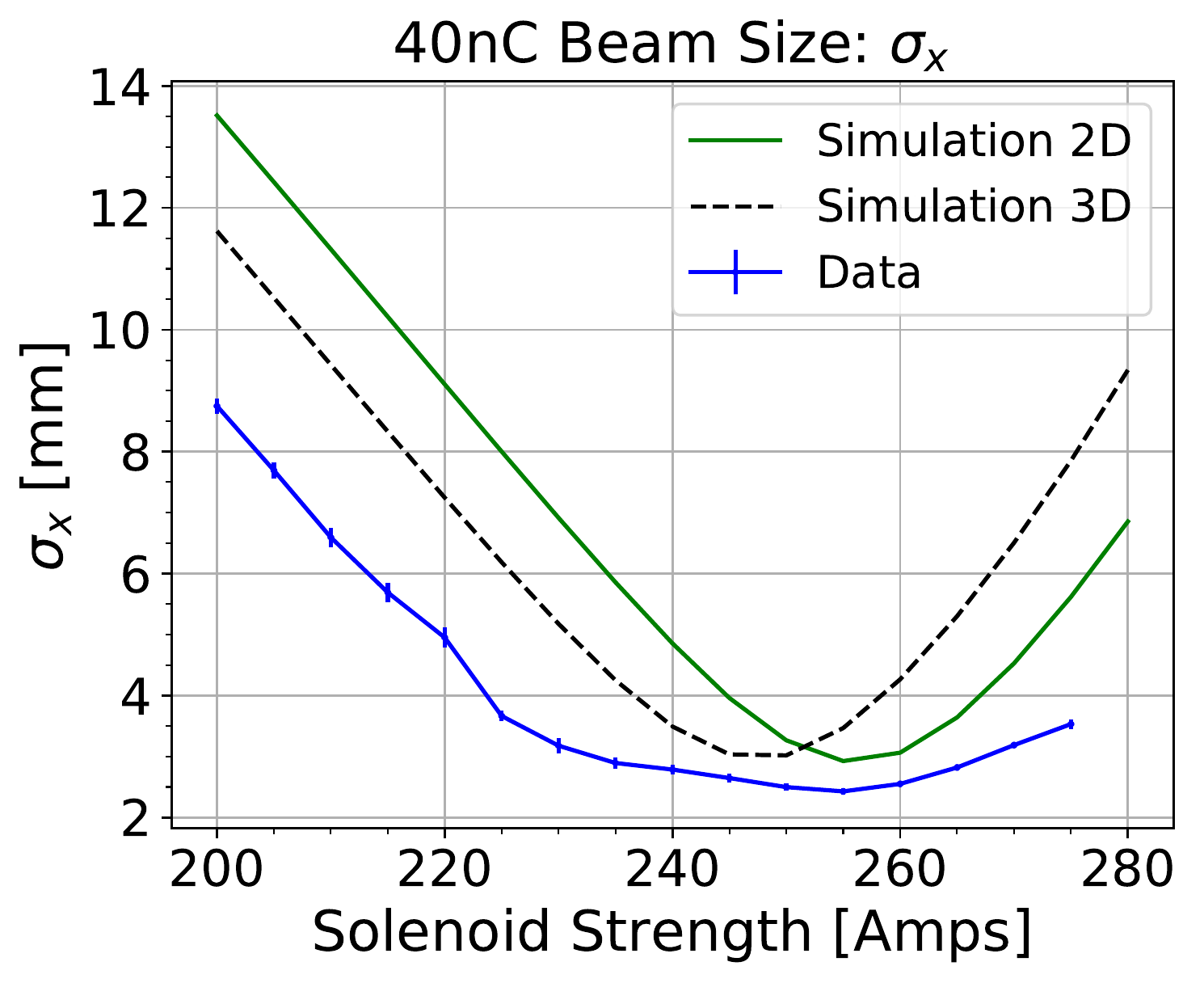}\includegraphics[width=0.45\linewidth]{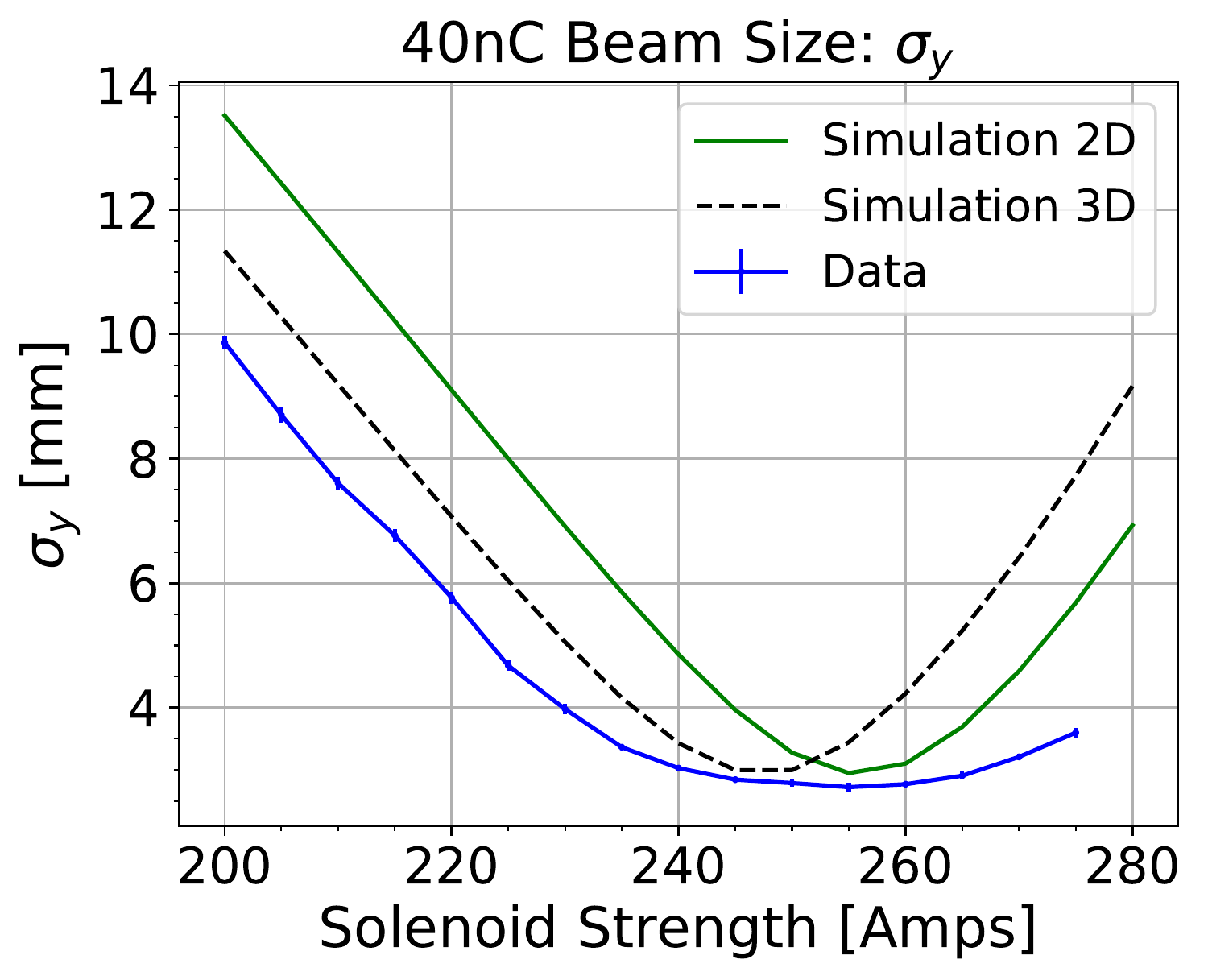}
    \caption{High charge solenoid scan at the AWA.}
    \label{fig:awasol2}
\end{figure}

First simulations of the gun at low and high charge were done.
The benchmark shown in Section~\ref{ssec:sc} is an example of this work.
In addition, a solenoid scan was performed and compared to measurements~\cite{neveu}, 
as shown in Figs.~\ref{fig:awasol2}.

While agreement is not exact, these plots were an important result for the AWA.
The gradient in \opalt was adjusted and several scans were done to reach the
comparison shown, i.e. initial simulations showed larger disagreement. 
The results indicated the gradient in the gun 
is lower than previously expected. 
This information helped inform future experiments 
and expectations at the AWA.

\subsubsection{Kicker Design}
\begin{figure}
    \centering
    \includegraphics[width=0.8\linewidth]{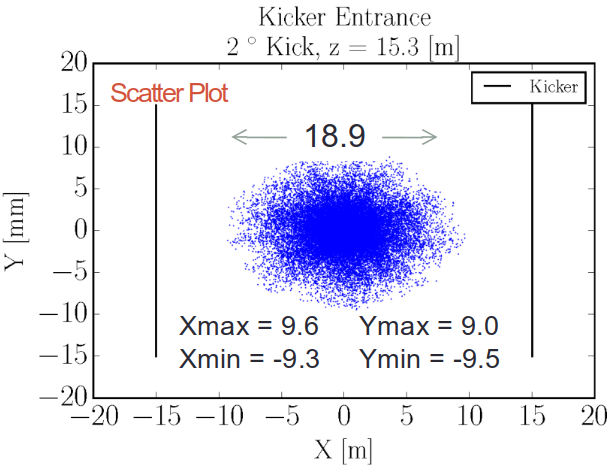}
    \caption{Estimated beam distribution at the entrance of the AWA kicker.}
    \label{fig:kickerscatter}
\end{figure}

\opalt was used heavily to help design a high charge kicker for the AWA.
Pre-fabrication, estimates of the beam size were simulated at \SI{40}{nC}, 
to determine how large the gap between the kicker plates needed to be, 
see Fig.~\ref{fig:kickerscatter} for an example of this type of simulation.

\begin{figure}
    \centering
    \includegraphics[width=0.85\linewidth]{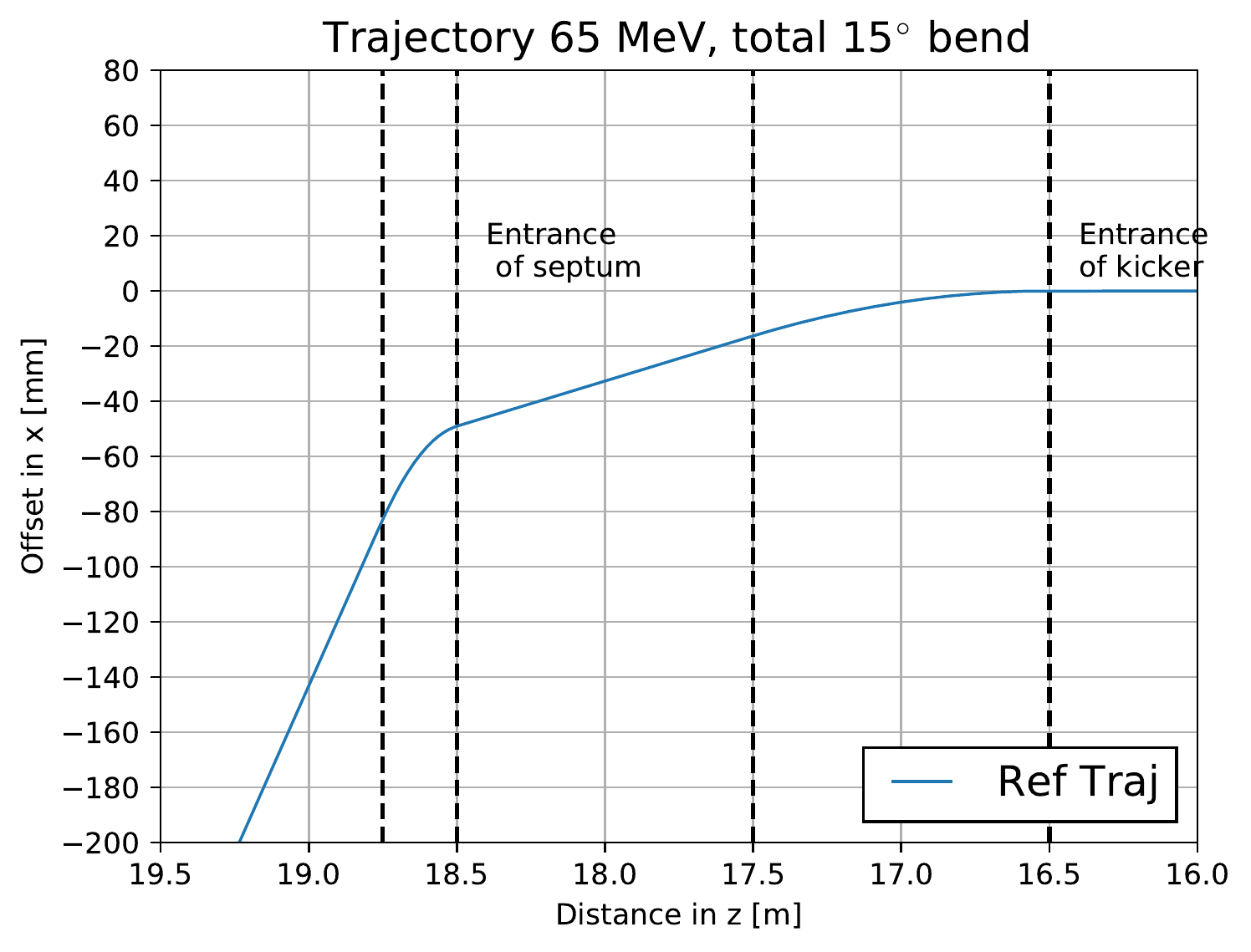}
    \caption{Transverse trajectory of the bunch as it travels through a kicker
    and septum. Note the beam is traveling from right to left, 
    as that is the orientation in the AWA tunnel.}
    \label{fig:traj}
\end{figure}

Another design consideration was the spacing between the 
kicker and subsequent septum that would be installed.
This depended heavily on the angle provided by the kicker, 
and the resulting transverse offset of the beam.
The ability to place elements in a 3D coordinate system 
(in \opalt 2.0), was crucial in providing realistic estimates on 
the transverse offset a kicker could provide to the beam centroid. 
These results, see Fig.~\ref{fig:traj}, showed 
a transverse offset of \SI{40}{mm} or more was possible \SI{1}{m}
downstream of the kicker. 
The results in Fig.~\ref{fig:traj}, along with the like shown 
Fig.~\ref{fig:kickerscatter} were critical in 
determining the length and angle of the kicker before fabrication. 

\subsubsection{Optimization}
Several multi-objective optimizations of the 
high charge beam line at AWA were performed 
using the built in genetic algorithm (GA) in \opalt. 
The objective was to optimize the 3D beam size, emittance, and energy spread 
leading into future TBA sections.
Constraints were also used to ensure beam sizes near the kicker
did not exceed or approach the gap width.
Details regarding this work can be found in~\cite{neveu,Neveu:2013ues}.

\subsection{The DAE$\delta$ALUS \& IsoDAR machines} \label{ssec:benchdaed}
\begin{figure}[t]
\centering
	\includegraphics[width=1.0\columnwidth]
	                {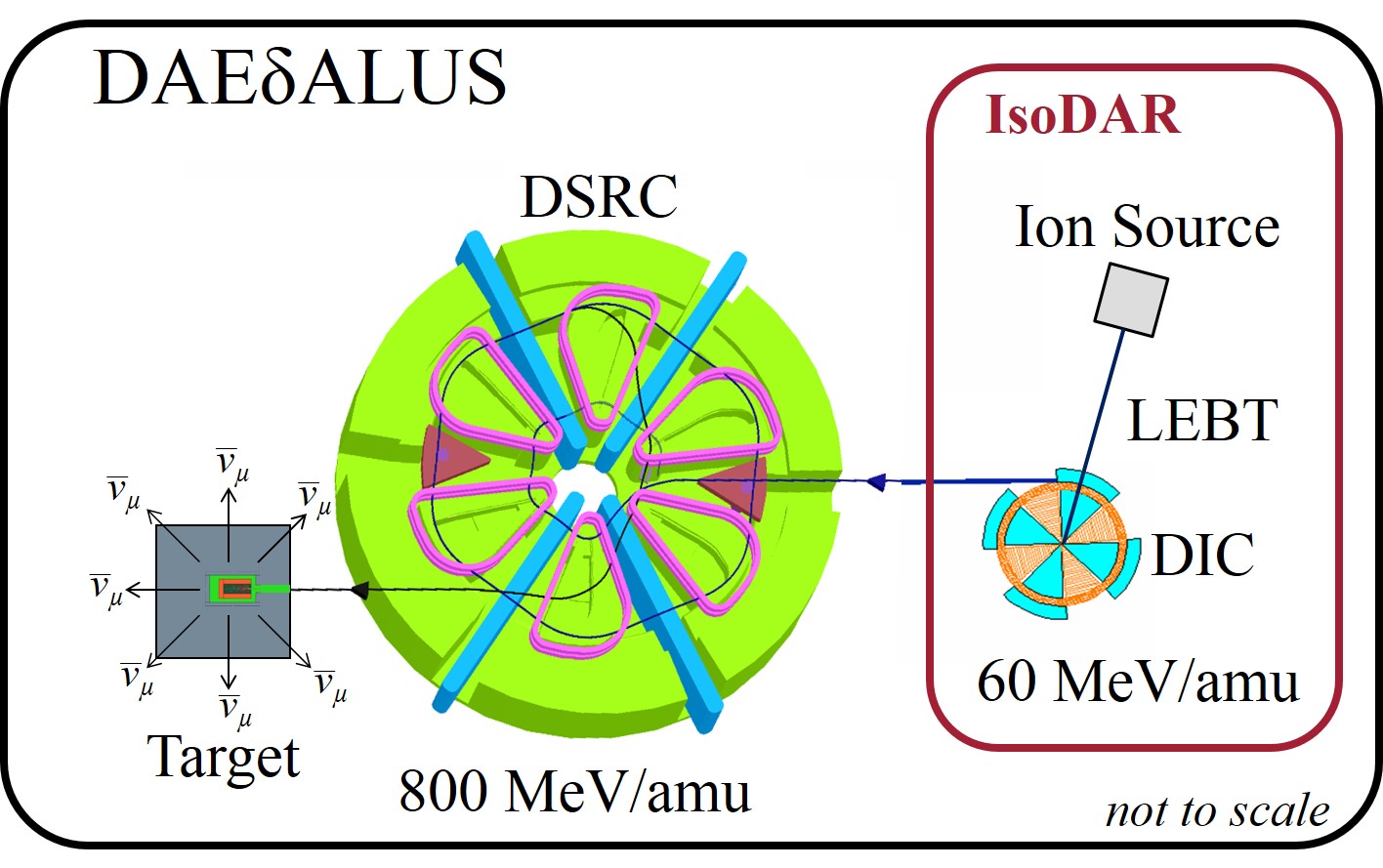}
	\caption{Schematic of the \DD facility. \htp is produced in the ion source,
	         transported to the \DD Injector Cyclotrons (DIC) and accelerated 
	         to 60 MeV/amu. Ions are subsequently extracted from the cyclotron
	         and injected into the \DD Superconducting Ring Cyclotron (DSRC),
	         where they are accelerated to 800 MeV/amu. During the highly 
	         efficient stripping extraction, 5 emA of \htp becomes 10 emA of
	         protons which impinge on the neutrino production target.}
	\label{fig:daedalus_cartoon}
\end{figure}

The Decay At-rest Experiment for $\delta_{CP}$ studies At a 
Laboratory for Underground Science (\DD) \cite{abs:daedalus, aberle:daedalus} 
is a proposed experiment to measure CP violation in the neutrino sector.
A schematic view of one \DD complex is shown in \figref{fig:daedalus_cartoon}. 
\htp is produced in an ion source, transported to the \DD Injector Cyclotrons 
(DIC), and accelerated to 60 MeV/amu. The reason for using \htp instead of
protons is to overcome space charge limitations of the high required beam
intensity of 10 emA of protons on target. \htp gives 2 protons for each 
unit of charge transported, thus mitigating the risk.
The ions are subsequently extracted from the cyclotron and injected into the 
\DD Superconducting Ring Cyclotron (DSRC) where they are accelerated to 
800 MeV/amu. During the highly efficient stripping extraction, the 5 emA of 
\htp become 10 emA of protons which impinge on the neutrino production target
(carbon) producing a neutrino beam virtually devoid of \nuebar.
In a large detector, one can then look for \nuebar appearance through 
neutrino oscillations.
As is depicted in \figref{fig:daedalus_cartoon}, the injector stage of \DD
can be used for another experiment: 
The Isotope Decay At Rest experiment IsoDAR \cite{adelmann:isodar, bungau:isodar}.
In IsoDAR, the 60 MeV/amu \htp will impinge on a beryllium target creating 
a high neutron flux. The neutrons are captured on $^7$Li surrounding the target.
The resulting $^8$Li beta-decays producing a very pure, isotropic \nuebar beam
which can be used for \nuebar disappearance experiments. IsoDAR is a definitive
search for so-called "sterile neutrinos", proposed new fundamental particles
that could explain anomalies seen in previous neutrino oscillation experiments.

At the moment, \opal is used for the simulation of three very important parts 
of the \DD and IsoDAR systems:
\begin{enumerate}
\item The spiral inflector (cf. \secref{sssec:benchinfl})
\item The \DD Injector Cyclotron (DIC), which is identical to the 
      IsoDAR cyclotron (cf. \secref{sssec:isodar}). 
\item The \DD Superconducting Ring Cyclotron (DSRC) for final acceleration 
      (cf. \secref{sssec:daedalus}). 
\end{enumerate}
The preliminary \opal simulation results of these three items will be briefly
discussed in the following subsections.

\subsubsection{Spiral inflector simulations \label{sssec:benchinfl}}
The goal of IsoDAR of accelerating 5 emA of \htp is ambitious and, due to controlled
beam losses in the central region, the current injected through the spiral inflector
can be a factor 5 higher than the extracted current, which leads to strong
space-charge forces and non-negligible mirror-charge effects. This challenge 
prompted the development of the new \spiral mode in \opalcycl reported on in
\cite{winklehner:spiral} and summarized in \secref{ssec:benchsc}, which lets the user 
provide \opal with an electric field map for the spiral electrodes and a geometry 
(cf. \secref{ssec:geom}), and, using the SAAMG field solver include this geometry
as boundary condition for the solver and for particle termination. As was described in
\secref{ssec:benchsc}, the agreement between experimental studies using an early iteration
of the IsoDAR spiral inflector and \opal was good. The latest development on 
spiral inflector design and simulation is a self-contained python program, first
reported on in \cite{weigel:spiral}, that allows 
for calculation of the spiral inflector shape from input parameters like beam energy and
electrode voltage, as well as field calculations using BEMPP \cite{smigaj:bempp}
and FEniCS \cite{alnaes:fenics}.
Exporting geometry- and field files for direct use in \opal are
a work in progress and an updated publication is forthcoming.

\subsubsection{The \DD Injector Cyclotron (DIC) / IsoDAR \label{sssec:isodar}}
\begin{figure}[t]
\centering
	\includegraphics[width=1.0\columnwidth]
	                {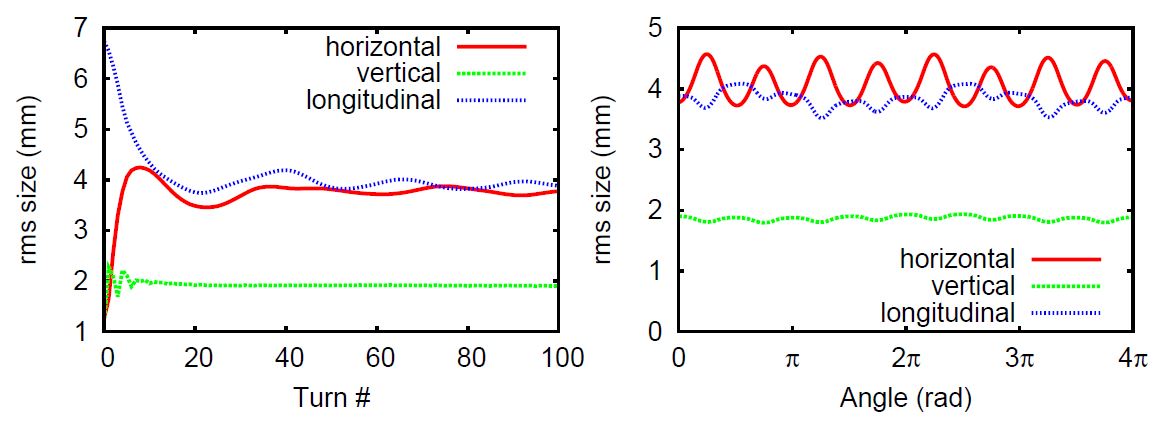}
	\caption{Left: RMS beam sizes as a function of turn number. 
	         Right: RMS beam sizes along the azimuth of the last turn. 
	         From \cite{yang:daedalus}.}
	\label{fig:isodar_stationary}
\end{figure}

Precise beam dynamics using high particle statistics and a method that can account
for non-linear space charge effects (particle-in-cell) is of the utmost importance 
for IsoDAR and the DIC.
This is due to the fact that so-called \emph{vortex motion} enables clean separation
of the final turns to do single turn extraction using an electrostatic septum. But only if
the beam exhibits this behaviour in a controlled manner.
This effect, the curling up of the beam in the longitudinal-radial plane (see also \cite{Bau:2011}), 
was first observed at the PSI Injector II cyclotron and successfully reproduced in \opalcycl \cite{yang:vortex}.
It is the reason beams of up to 2.7 emA have been extracted from Injector II. For IsoDAR and 
the DIC, vortex motion is taken into account already in the design phase, to enable a total
extractable beam current of 5 emA of \htp. First results were presented in 2013 \cite{yang:daedalus},
starting the tracking at an energy of 1.5 MeV/amu inside the cyclotron. In \cite{yang:daedalus}, Yang 
showed that even with an initially mismatched beam, a stationary, almost round distribution forms
after less than 10 turns. This can be seen in \figref{fig:isodar_stationary}, where the rms beam 
sizes are plotted versus turn number and along the azimuth of the final turn. 
By placing collimators during the first several
turns, the remaining halo could be cleaned up to the point where less than 150 W of beam were 
lost on the septum in the final turn. Since then, the tracking has been extended down to a starting 
energy of 194 keV/amu - which corresponds to a particle having completed the first turn after the
spiral inflector - with similar results \cite{jonnerby_central_2016}. However, the amount of beam that
has to be removed early on increased in these latter simulation, so that a starting current of 
6.65 emA was necessary to obtain an extractable beam current of 5.1 emA. Further refinement of
the solution is ongoing at MIT.

\subsubsection{The \DD Superconducting Ring Cyclotron (DSRC) \label{sssec:daedalus}}
The beam dynamics of the DSRC were discussed in \cite{yang:daedalus}. The usage of \htp
in the acceleration process first in the DIC/IsoDAR cyclotron and then in the 
DSRC, has the advantage of being able to strip away the loosely bound electron with a
\emph{stripper foil}, thereby freeing the two protons originally bound in the \htp molecular ion.
These protons have different magnetic rigidity than \htp and it is possible to devise a path
through the cyclotron for clean extraction (close to 100\% efficient).
In \cite{yang:daedalus}, the authors identify two major pitfalls. One, the vertical focusing 
becomes weak for higher turn numbers (220 and up). Two, stripping extraction usually instroduces
energy spread, because ions are extracted from several different turn numbers simultaneously.
Through careful simulations using \opal, they determined that, while indeed the vertical beam size 
will grow significantly after turn 220 (up to 30 mm), the vertical gap of the cyclotron is able to accommodate this increase. In addition, \opalcycl was updated during that study to 
simulate several neighboring bunches, to include the space-charge coupling of the close, sometimes overlapping turns. A simple stripping process (100\% efficiency, no straggling, no additional interactions beyond removing the electron) was simulated, using these overlapping turns of
\htp bunches, and the resulting protons were tracked through the cyclotron to be extracted. 
Again, the high-statistics \opal simulations showed that while energy spread and beam growth were present, they are not prohibitively large.


%

\section*{Acknowledgements}
The contributions of various individuals that had considerable influence on the
development of \opal\ are acknowledged, namely Chris Iselin, John Jowett, Julian Cummings, Ji Qiang, Robert Ryne and Stefan Adam. For the H5root  analysis tool credits go to Thomas Schietinger, for the initial development. Valeria Rizzoglio was benchmarking the particle matter interaction model.

For the work related to AWA, we gratefully acknowledge the computing resources provided on Bebop, a high-performance computing cluster operated by the Laboratory Computing Resource Center at Argonne National Laboratory

\section{References}
\bibliography{biblio}

\newpage
\appendix
\section{Materials with their Parameters}
In order to simulate not only the degrader but also foils and collimator in the beam transport line and the nozzle at the isocenter of the gantries, new materials have been implemented in OPAL as listed in \tabref{table:Materials}.
\begin{table*}[htb] \small
	\centering
 	\begin{tabular}{|c|c|c|c|c|c|c|c|c|}
	\hline
	\textbf{Material} 	& \textbf{Z}	&	\textbf{A}	 	&	\textbf{$\rho$}  [$g/cm^3$]	 &	\textbf{X0}  	[$g/cm^2$]	 &	\textbf{A2}	&	\textbf{A3}	&	\textbf{A4}	&	\textbf{A5}\\
	\hline
	Aluminum		&	13		&   	   26.98		&	          2.7		&	24.01		&	4.739		&	2766		&	164.5		&	2.023E-02\\
	\hline
	Copper			&	29		&	  63.54		&		8.96		&	12.86		& 	4.194		&	4649		&	81.13		&	2.242E-02\\
	\hline
	Graphite			&	6		&	   12			&	       2.210	&	42.7			&	2.601		&	1701		&	1279		&	1.638E-02\\
	\hline
	GraphiteR6710		&	6		&	   12			&	       1.88		&	42.7			&	2.601		&	1701		&	1279		&	1.638E-02\\
	\hline
	Titan			&	22		&	  47.8		&	      4.54		&	16.16		&	5.489		&	5260		&	651.1		&	8.930E-03\\
	\hline
	Alumina			&	50		&	101.96		&	      3.97		&	27.94		&	7.227		&	11210		&	386.4		&	4.474E-03\\
	\hline
	Air				&	7		&	14			&	    0.0012		&	37.99		&	3.350		&	1683		&	1900		&	2.513E-02\\
	\hline
	Kapton			&	6		&	12			&	1.4			&	39.95		&	2.601		&	1701		&	1279		&	1.638E-02\\
	\hline
	Gold				&	79		&	197			&	19.3			&	6.46			&	5.458		&	7852		&	975.8		&	2.077E-02\\
	\hline
	Water			&	10		&	18			&	1			&	36.08		&	2.199		&	2393		&	2699		&	1.568E-02\\
	\hline
	Mylar			&	6.702	&	12.88		&	1.4			&	39.95		&	3.350		&	1683		&	1900		& 	2.5133E-02\\
	\hline
	Beryllium		 	&	4		&	9.012		&	1.848		&	65.19		&	2.590		&	966.0		&	153.8		& 	3.475E-02\\
	\hline
	Mo				&	4.68		&	95.94		&	10.22		&	9.8			&	7.248		&	9545		&	480.2		& 	5.376E-03\\
	\hline
	\end{tabular}
		\caption{List of materials with their parameters implemented in OPAL}
		\label{table:Materials}
\end{table*}

The $A_i$ coefficients (i = 2...5) for the energy loss calculation at low energy have been found on ICRU \cite{ICRU} (Table 3.1).

\end{document}